\documentclass[twocolumn]{aastex631}

\usepackage{mathrsfs}

\usepackage{rotating}
\usepackage{ltablex}
\usepackage{booktabs}

\newcommand{\rev}[1]{\textcolor{black}{#1}}

\newcommand{\code}[1]{{\texttt{#1}}}

\newcommand{\kepler}{\textit{Kepler}\ }
\newcommand{\tess}{\textit{TESS}\ }

\begin{document}

\title{A transiting super-Earth in the radius valley and an outer planet candidate around HD 307842\footnote{This paper includes data gathered with the 6.5 meter Magellan Telescopes located at Las Campanas Observatory, Chile.}}

\correspondingauthor{Xinyan Hua, Sharon Xuesong Wang}
\email{huaxinyan1996@gmail.com, sharonw@tsinghua.edu.cn}

\author[0000-0001-7916-4371]{Xinyan Hua}
\affiliation{Department of Astronomy, Tsinghua University, Beijing 100084, China}

\author[0000-0002-6937-9034]{Sharon Xuesong Wang} 
\affiliation{Department of Astronomy, Tsinghua University, Beijing 100084, China}

\author{Johanna K. Teske}  
\affiliation{Earth and Planets Laboratory, Carnegie Institution for Science, 5241 Broad Branch Road, NW, Washington, DC 20015, USA}

\author{Tianjun Gan} 
\affiliation{Department of Astronomy, Tsinghua University, Beijing 100084, China}

\author[0000-0002-1836-3120]{Avi Shporer} 
\affiliation{Department of Physics and Kavli Institute for Astrophysics and Space Research, Massachusetts Institute of Technology, Cambridge, MA 02139, USA}

\author[0000-0002-4891-3517]{George Zhou} 
\affiliation{Centre for Astrophysics, University of Southern Queensland, Queensland, Australia}

\author[0000-0002-3481-9052]{Keivan G.\ Stassun}
\affiliation{Department of Physics and Astronomy, Vanderbilt University, Nashville, TN 37235, USA}


\author[0000-0003-2935-7196]{Markus Rabus} 
\affiliation{Departamento de Matem{\'a}tica y F{\'i}sica Aplicadas, Facultad de Ingenier{\'i}a, Universidad Cat{\'o}lica de la Sant{\'i}sima Concepci{\'o}n, Alonso de Rivera 2850, Concepci{\'o}n, Chile}

\author[0000-0002-2532-2853]{Steve~B.~Howell} 
\affiliation{NASA Ames Research Center, Moffett Field, CA 94035, USA}

\author{Carl Ziegler} 
\affiliation{Department of Physics, Engineering and Astronomy, Stephen F. Austin State University, 1936 North St, Nacogdoches, TX 75962, USA}

\author[0000-0001-6513-1659]{Jack J. Lissauer}
\affiliation{Space Science \& Astrobiology Division, MS 245-3, NASA Ames Research Center, Moffett Field, CA 94035, USA}

\author[0000-0002-4265-047X]{Joshua~N.~Winn}
\affiliation{Department of Astrophysical Sciences, Princeton University, 4 Ivy Lane, Princeton, NJ 08544, USA}

\author[0000-0002-4715-9460]{Jon~M.~Jenkins}
\affiliation{NASA Ames Research Center, Moffett Field, CA 94035, USA}

\author[0000-0002-8219-9505]{Eric B. Ting} 
\affiliation{NASA Ames Research Center, Moffett Field, CA 94035, USA}

\author{Karen A. Collins} 
\affiliation{Center for Astrophysics \textbar \ Harvard \& Smithsonian, 60 Garden Street, Cambridge, MA 02138, USA}

\author[0000-0003-3654-1602]{Andrew W. Mann} 
\affiliation{Department of Physics and Astronomy, The University of North Carolina at Chapel Hill, Chapel Hill, NC 27599-3255, USA}

\author[0000-0003-4027-4711]{Wei Zhu} 
\affiliation{Department of Astronomy, Tsinghua University, Beijing 100084, China}

\author{Su Wang}
\affiliation{Purple Mountain Observatory, No.10 Yuanhua Road, Nanjing, 210023, China}

\author[0000-0003-1305-3761]{R. Paul Butler}
\affiliation{Earth and Planets Laboratory, Carnegie Institution for Science, 5241 Broad Branch Road, NW, Washington, DC 20015, USA}

\author[0000-0002-5226-787X]{Jeffrey D. Crane}
\affil{Observatories of the Carnegie Institution for Science, 813 Santa Barbara Street, Pasadena, CA 91101, USA}

\author{Stephen A. Shectman}
\affil{Observatories of the Carnegie Institution for Science, 813 Santa Barbara Street, Pasadena, CA 91101, USA}




\author[0000-0002-0514-5538]{Luke~G.~Bouma} 
\affiliation{Cahill Center for Astrophysics, California Institute of Technology, Pasadena, CA 91125, USA}

\author{C\'{e}sar Brice\~{n}o} 
\affiliation{Cerro Tololo Inter-American Observatory, Casilla 603, La Serena, Chile}

\author[0000-0003-2313-467X]{Diana~Dragomir} 
\affiliation{Department of Physics and Astronomy, University of New Mexico, 210 Yale Blvd NE, Albuquerque, NM 87106, USA}

\author{William~Fong} 
\affiliation{Department of Physics and Kavli Institute for Astrophysics and Space Research, Massachusetts Institute of Technology, Cambridge, MA 02139, USA}

\author{Nicholas Law} 
\affiliation{Department of Physics and Astronomy, The University of North Carolina at Chapel Hill, Chapel Hill, NC 27599-3255, USA}

\author{Jennifer V.~Medina} 
\affiliation{Space Telescope Science Institute, 3700 San Martin Drive, Baltimore, MD, 21218, USA}

\author{Samuel N.~Quinn}
\affiliation{Center for Astrophysics \textbar \ Harvard \& Smithsonian, 60 Garden Street, Cambridge, MA 02138, USA}

\author{George~R.~Ricker}
\affiliation{Department of Physics and Kavli Institute for Astrophysics and Space Research, Massachusetts Institute of Technology, Cambridge, MA 02139, USA}

\author[0000-0001-8227-1020]{Richard P. Schwarz} 
\affiliation{Center for Astrophysics \textbar \ Harvard \& Smithsonian, 60 Garden Street, Cambridge, MA 02138, USA}

\author[0000-0002-6892-6948]{Sara~Seager}
\affil{Department of Earth, Atmospheric, and Planetary Sciences, Massachusetts Institute of Technology, Cambridge, MA 02139, USA}
\affil{Department of Physics and Kavli Institute for Astrophysics and Space Research, Massachusetts Institute of Technology, Cambridge, MA 02139, USA}
\affil{Department of Aeronautics and Astronautics, Massachusetts Institute of Technology, Cambridge, MA 02139, USA}

\author[0000-0003-3904-6754]{Ramotholo Sefako} 
\affiliation{South African Astronomical Observatory, \\
P.O. Box 9, Observatory, Cape Town 7935, South Africa}

\author[0000-0003-2163-1437]{Chris Stockdale} 
\affiliation{Hazelwood Observatory, Australia}

\author{Roland~Vanderspek}
\affiliation{Department of Physics and Kavli Institute for Astrophysics and Space Research, Massachusetts Institute of Technology, Cambridge, MA 02139, USA}

\author{Joel Villase{\~ n}or} 
\affiliation{Department of Physics and Kavli Institute for Astrophysics and Space Research, Massachusetts Institute of Technology, Cambridge, MA 02139, USA}




\begin{abstract}

We report the confirmation of a \textit{TESS}-discovered transiting super-Earth planet orbiting a mid-G star, HD 307842 (TOI-784). The planet has a period of 2.8 days, and the radial velocity (RV) measurements constrain the mass to be $9.67^{+0.83}_{-0.82}\ \rm{M_{\oplus}}$.
We also report the discovery of an additional planet candidate on an outer orbit that is most likely non-transiting. The possible periods of the planet candidate are approximately 20 to 63 days, with the corresponding RV semi-amplitudes expected to range from 3.2 to 5.4 m/s and minimum masses from $12.6$ to $31.1\ \rm{M_{\oplus}}$.
The radius of the transiting planet (planet b) is $1.93^{+0.11}_{-0.09}\ \rm{R_{\oplus}}$, which results in a mean density of $7.4^{+1.4}_{-1.2}\ \rm{g/cm^3}$ suggesting that TOI-784b is likely to be a rocky planet though it has a comparable radius to a sub-Neptune. We found TOI-784b is located at the lower edge of the so-called ``radius valley'' in the radius vs. insolation plane, which is consistent with the photoevaporation or core-powered mass loss prediction. 
The \tess data did not reveal any significant transit signal of the planet candidate, and our analysis shows that the orbital inclinations of planet b and the planet candidate are ${88.60^{\circ}}^{+0.84}_{-0.86}$ and $\leqslant 88.3^{\circ}-89.2^{\circ}$, respectively. More RV observations are needed to determine the period and mass of the second object, and search for additional planets in this system.

\end{abstract}



\section{Introduction}
\label{sec:introduction}
 
Planets in extrasolar systems are common in the universe. In the past few decades, over 5000 exoplanets were discovered and confirmed. NASA's \kepler space telescope discovered thousands of transiting planet candidates with most of them having a size between Earth and Neptune \citep{borucki2010}. \citet{Fulton2017} studied the size distribution of 2025 \kepler planets (the California-\kepler Survey sample) in fine detail and found a bimodal structure with a gap near $1.5-2.0\ \rm{R_{\oplus}}$ separating super-Earths and sub-Neptunes, which is referred as the ``radius valley''. \citet{Weiss2018b} further claimed that this valley also exists in multi-planet systems. 

The formation process of the radius valley and its position as a function of other planetary or host stars' parameters are still under debate. A positive relationship between the transition radius and planet insolation flux was derived from the CKS sample \citep{Martinez2019, Petigura2022}, which can be explained by the photoevaporation scenario \citep[e.g.][]{owen_wu2017, Lopez&Rice2018, Fulton2018}, where the close-in planets become stripped cores while the outer ones can keep their gas envelopes and thus have larger sizes. On the other hand, \citet{Cloutier&Menou2020} estimated an opposite correlation using 275 \kepler and 53 \textit{K2} planets around M and K type stars, which is consistent with a gas-poor formation theory \citep[e.g.,][]{Lopez&Rice2018}, where most rocky planets are formed after the discs dissipate and thus without initial gaseous envelopes. 

Radial velocity (RV) follow-up observations on planets with sizes near the radius valley could shine more light on the structure and origin of the valley, especially for the ``keystone" planets where different models predict different fractions of volatile (e.g., \citealt{Cloutier&Menou2020,Cloutier2020}). The mass measurements from RVs would provide some constraints on the bulk composition of transiting planets, thus revealing how planetary composition might shift across the radius valley (e.g., \citealt{Luque2022}). Obtaining precise mass measurements of super-Earths and sub-Neptunes is also important for reliably characterizing the composition of their atmosphere (if any; \citealt{batalha2019}).

In addition to the mass and bulk density measurement, RV follow-up can also reveal additional planets in the system, either non-transiting or yet to be detected by transits, which is more important for mapping out the system architecture for transiting planets discovered by the \textit{Transiting Exoplanet Survey Satellite} (\textit{TESS}; \citealt{ricker2015}) given its relatively shorter baseline compared with \textit{Kepler}. Studies on planet multiplicity found that multi-planet systems are common (e.g., 40\% among systems with super-Earths or sub-Neptunes discovered by \textit{Kepler}; \citealt{batalha2013}), so RV follow-up observations on transiting planets often turn out to be fruitful \citep[e.g.,][]{Lacedelli2021,Lubin2022}. Revealing the non-transiting planets in systems with single transiting planets can be of particular interest, as it directly addresses the previously identified ``\kepler dichotomy", where a surplus of \kepler single-transiting systems was inconsistent with predictions from earlier population synthesis models with relatively low mutual inclinations, but consistent with an additional population of multi-planet systems with large mutual inclinations \citep[e.g.,][]{lissauer2011,hansen2013,ballardjohnson2016}. This ``dichotomy" appears to hold for both M dwarfs and Sun-like stars, and it can not be fully explained by selection biases \citep[e.g.][]{Zink2019}. Different scenarios were proposed to explain the \kepler dichotomy \citep[e.g.][]{Mulders2018, zhu2018, He2019, He2020}, but the intrinsic mutual inclination distribution for multi-planet systems still remains uncertain. More samples of well-charactered multi-planet systems, either with RVs or TTVs, are needed for investigations on the intrinsic mutual inclination distribution of planets and studies on system architectures in general (e.g., see review by \citealt{Weiss2022}).

In this work, we report a \tess discovered transiting super-Earth, TOI-784b, and an additional Neptune-mass planet candidate in the system using RV follow-up data taken by the Magellan \tess Survey (MTS; \citealt{Teske2021}). We organize this paper as follows: we first describe our observational data and data reduction processes in Section~\ref{sec:observations}, as well as results from direct imaging. We then introduce the stellar characterization in Section~\ref{sec:stellar-characterisation}. Section~\ref{sec:planetary-analysis} presents the detailed photometry and radial velocity modeling to constrain the planetary properties. We discuss our results and present our conclusions in Section~\ref{sec:discussion}.

\section{Observations}
\label{sec:observations}

\subsection{Photometry}
\label{sec:Photometry} 




\subsubsection{TESS}
\label{sec:TESS}

The \tess mission is an all-sky survey to discover transiting exoplanets \citep{ricker2015}, with a field of view of $24^{\circ}\times 96^{\circ}$, observing in Sectors each lasting about 27 days. From July 2018 to July 2020, \tess mapped almost the whole sky in its two-year Primary Mission and then re-observed the southern ecliptic hemisphere during Year 3 (July 2020-July 2021) for its first Extended Mission. \tess recently completed its first extended mission slightly over four years from the start of science observations in July 2018 and has now commenced its second extended mission. 

HD 307842 (TIC 460984940, hereafter TOI-784) was observed by the \tess mission and announced as TOI-784 after its Cycle 1 observation in Sectors 10 (March 26 to April 22, 2019) and 11 (April 22 to May 21, 2019) with a two-minute cadence using camera 3. In Cycle 3, TOI-784 was revisited with the same mode in Sector 37 (April 2 to April 28, 2021) and 38 (April 28 to May 26, 2021). The raw photometric data were first reduced by the Science Processing Operations Center \citep[SPOC;][]{Jenkins:2016} pipeline at NASA Ames Research Center, which extracted the light curve using Simple Aperture Photometry (SAP) and further calibrated for instrument systematics using the Presearch Data Conditioning (PDC) algorithm. 

The SPOC Transiting Planet Search \citep[TPS;][]{Jenkins2002, Jenkins2010, 2020TPSkdph} of the light curve via an adaptive, noise-compensating matched filter revealed a transit signal with a 2.8-day period on 23 May 2019. An initial limb-darkened transit model was fitted \citep{Li:DVmodelFit2019} and a suite of diagnostic tests were conducted to help make or break the planetary nature of the signal(s) \citep{Twicken:DVdiagnostics2018}, presented in the Data Validation reports available on the Mikulski Archive for Space Telescopes\footnote{\url{https://archive.stsci.edu}} (MAST). In this case the transit signal failed the ghost diagnostic test, but visual inspection of the difference images indicates that the transit source is located on the target. The \tess Science Office reviewed the vetting reports and issued an alert for TOI-784b on 5 June 2019 \citep{guerrero:TOIs2021ApJS}. 

The Presearch Data Conditioning Simple Aperture Photometry light curve file \citep[PDCSAP flux;][]{Stumpe2012, Smith2012, Stumpe2014} for photometry analyses in this work (see Section~\ref{sec:photometry-analysis}) was downloaded from MAST using the \code{lightkurve} package \citep{Lightkurve}.

\subsubsection{Las Cumbres Observatory: Sinistro}
\label{sec:LCO-Sinistro}

We collected three sets of ground-based light curves for TOI-784 using the Las Cumbres Observatory Global Telescope\ (LCOGT\footnote{\url{https://lco.global/}}) network \citep{Brown2013} on 26$^{\rm th}$ Feb. 2020, 13$^{\rm rd}$ Mar. 2020 and 21$^{\rm st}$ Jan. 2021 to refine the transit ephemeris and rule out the nearby eclipsing binary scenario. We used the {\tt TESS Transit Finder}, which is a customized version of the {\tt Tapir} software package \citep{Jensen2013}, to schedule the transit observation. All three observations were taken with the Sinistro cameras in the Pan-STARRS $z$-short band ($z_{s}$) with exposure times of 70, 70, and 30 s. We carried out a photometric analysis and extracted the light curves using {\tt AstroImageJ} \citep{Collins2017}. We excluded all nearby stars within $2.5\arcmin$ as the source that may cause the \tess signal with brightness difference down to $ \Delta T \sim 9.5 $ mag and tentatively detected the signal on target. The data are publicly available on ExoFOP\footnote{\url{https://exofop.ipac.caltech.edu/tess/target.php?id=88902249}}.

\subsubsection{Other Archival Ground-based Photometry}
\label{sec:archival-Ground-based-Photometry}

\begin{figure}
	\includegraphics[width=\columnwidth]{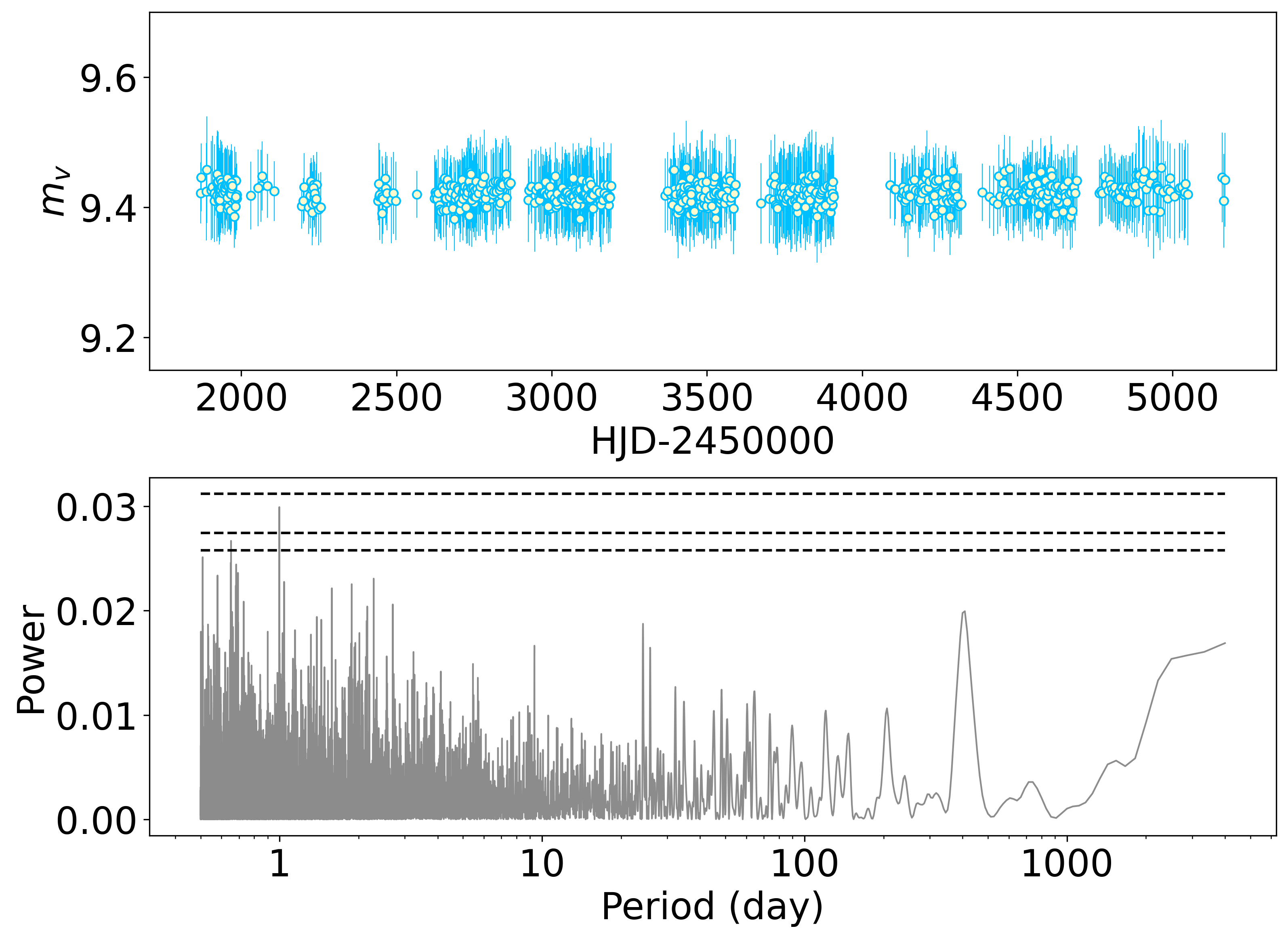}
    \caption{The ASAS V-band photometry (top) and its GLS periodogram. The three dashed lines from top to bottom in the lower panel correspond to 1\%, 5\%, and 10\% significance threshold levels, respectively. The maximum signal appears at around 1 day, which is likely an alias due to the observing cadence. No significant signal was identified for the stellar rotation.}
    \label{fig:784gls}
\end{figure}

We found 963 $V$-band archival measurements of TOI-784 from the ASAS database\footnote{\url{http://www.astrouw.edu.pl/asas/?page=aasc&catsrc=asas3}} spanning from November 21, 2000, to December 3, 2009.\footnote{We also found archival photometry data from ASAS-SN (\url{https://asas-sn.osu.edu/photometry}), in which we saw a long term trend spanning in the time baseline ($\sim 904$~days), so we did not adopt it. We found no archival data from WASP (\url{https://wasp.cerit-sc.cz/form}).} Following the ASAS documentation, we selected 883 data points ranked as grades A and B, which represent the highest-quality measurements. We applied a 3-sigma clipping rejection, after which 35 measurements were masked and 848 were left. Considering the $V$ magnitude of our target ($\sim 9.4$) and the observing aperture scale, the data points marked as ``$\rm{MAG\_3}$'' were used in our analysis as recommended by ASAS\footnote{Indices from 0-4 indicate 5 apertures of 2-6 pixels wide. Small ones are better for faint stars while large ones for bright stars, roughly following the relation of $aperture\ index = 12-V$. See \url{http://www.astrouw.edu.pl/asas/explanations.html} for more detailed explanations.}. We calculated \rev{the Generalized Lomb-Scargle (GLS) Periodogram} using the python module \texttt{PyAstronomy.pyTiming.pyPeriod.Gls()} to search for the rotation signal of the host star\footnote{A same analysis was applied using the \tess light curve, which alerted no evident signal as expected.}. Results are shown in Figure~\ref{fig:784gls}; the three black dashed lines from top to bottom in the figure correspond to false-alarm probabilities ($\rm{FAP}$) of 1\%, 5\%, and 10\%, \rev{which means that there is a 1\%, 5\%, or 10\% chance, respectively, that the observed signal is a false positive. The FAP values were computed via the default function within the \texttt{Gls()} module with the default normalization assumption described in \cite{GLS2009}.} We found no significant rotation modulation, suggesting that TOI-784 is probably an old, slowly rotating and photometrically quiet star. The peak around one day is likely an alias resulting from the observation cadence.

\subsection{Spectroscopy}
\label{sec:Spectroscopy}
\subsubsection{PFS} 
\label{sec:PFS}

We collected a total of 35 RV data points using the Planet Finder Spectrograph \citep[PFS;][]{crane2006, crane2008, crane2010} on the 6.5-meter Magellan II Clay telescope at Las Campanas Observatory in Chile. PFS is a high-resolution, optical echelle spectrograph that covers a wavelength band of 391--734~nm with a resolving power of R $\sim$ 130,000 using the $0.3\arcsec\times 2.5\arcsec$ slit. The RV precision of PFS on nearby, bright, and photospherically quiet stars is typically 0.5--1.0~m/s. The spectral data reduction and RV extraction were performed using a customized pipeline \citep{butler1996}.

Our PFS data were obtained as part of the Magellan \tess Survey \citep[MTS;][]{Teske2021}, a project designed to study a sample of 30 well-characterized small planets with radii $< 3R_{\oplus}$ to reveal their population statistics such as the M-R relation \citep{Wolfgang2016, Weiss2018b}, the ``radius gap'' around $1.8R_{\oplus}$ \citep{Fulton2017, Petigura2022}, and other stellar properties/system architecture in order to reveal their formation and evolution processes. A sample of 30 targets were selected using a quantified merit function among the \tess Objects of Interest (TOIs) from the Year 1 observations of \tess (therefore only includes the southern hemisphere), and TOI-784 was among the initial sample of 30 at the beginning of MTS but later on dropped off the list due to the refinement of stellar and planetary parameters of TOIs over time.


MTS adopts a specific observing strategy to minimize selection bias (see \citealt{Teske2021} for more details about the observing cadence design). TOI-784 received a High Cadence grade for its coverage, with 22 RV observations taken from UT March 6 to UT March 17 in 2020 and 9 RV data points between UT May 22 and UT May 29 in 2021, and MTS stopped observing TOI-784 afterward as it dropped out of the target list. We then collected four more data points from UT 2022-3-14 to 2022-3-25 in order to characterize the additional long-period planet candidate, as the RV fitting results using the first year's data showed a linear trend in the residuals (see Section~\ref{sec:RV} for more). 
With a typical exposure time of 10 to 20~minutes depending on the seeing, the reported internal RV precision $\sigma_{RV}$ of PFS on TOI-784 is 0.7--1.0~m/s. See Table~\ref{tab:rv-data-points} for all the PFS RVs used in this work.

\begin{tabularx}{\linewidth}{cccc}
\label{tab:rv-data-points} \\
\caption{RV measurements} \\
\toprule
Time & RV & Uncertainty & Instrument\\
 & (m/s) & (m/s) &  \\
\midrule
\endfirsthead
\toprule
Time & RV & Uncertainty & Instrument\\
 & (m/s) & (m/s) &  \\
\midrule
\endhead
\midrule
\multicolumn{1}{l}{\footnotesize(To be continued)}
\endfoot
\bottomrule
\endlastfoot
2458914.68102 & -6.02 & 0.83 & PFS \\
2458914.80040 & -8.24 & 0.87 & PFS \\
2458915.68459 & -3.32 & 0.79 & PFS \\
2458915.75693 & -3.82 & 0.78 & PFS \\
2458916.71986 & 0.34 & 0.77 & PFS \\
2458917.65647 & -6.89 & 0.74 & PFS \\
2458917.73457 & -5.19 & 0.74 & PFS \\
2458918.67521 & 0.27 & 0.91 & PFS \\
2458918.75249 & 2.39 & 0.72 & PFS \\
2458919.66161 & 0.12 & 0.68 & PFS \\
2458919.74236 & 3.05 & 0.70 & PFS \\
2458920.65817 & -5.21 & 0.64 & PFS \\
2458920.71749 & -4.78 & 0.64 & PFS \\
2458921.67109 & 3.56 & 0.65 & PFS \\
2458921.73069 & 6.03 & 0.62 & PFS \\
2458923.63457 & -0.01 & 0.71 & PFS \\
2458923.70633 & -1.72 & 0.75 & PFS \\
2458924.62670 & 6.63 & 0.61 & PFS \\
2458924.70608 & 6.77 & 0.70 & PFS \\
2458924.77554 & 8.45 & 0.72 & PFS \\
2458925.62948 & 1.55 & 0.67 & PFS \\
2458925.71178 & -0.09 & 0.72 & PFS \\
2459356.50045 & -6.68 & 0.99 & PFS \\
2459356.57079 & -6.90 & 1.06 & PFS \\
2459358.57198 & 0.90 & 1.09 & PFS \\
2459359.49081 & -6.39 & 1.11 & PFS \\
2459359.58574 & -8.32 & 1.03 & PFS \\
2459361.45964 & 0.80 & 1.03 & PFS \\
2459361.58363 & 0.59 & 1.14 & PFS \\
2459363.46677 & 3.25 & 1.08 & PFS \\
2459363.55417 & 5.02 & 1.04 & PFS \\
2459652.72417 & -0.57 & 0.73 & PFS \\
2459654.70344 & 2.93 & 0.70 & PFS \\
2459657.67591 & 0.00 & 0.71 & PFS \\
2459663.67738 & -1.96 & 1.00 & PFS \\
\hline
2459336.22205515 & 15163.8 & 10.5 & NRES1 \\
2459344.41512404 & 15157.3 & 9.1 & NRES1 \\
2459346.23537912 & 15136.2 & 9.2 & NRES1 \\
2459348.25530748 & 15198.2 & 9.0 & NRES1 \\
2459353.21427107 & 15118.7 & 9.4 & NRES1 \\
2459357.35987432 & 15162.1 & 11.0 & NRES1 \\
2459360.33375668 & 15153.9 & 9.8 & NRES1 \\
2459369.30694576 & 15142.5 & 8.7 & NRES1 \\
2459376.32741634 & 15122.2 & 8.9 & NRES1 \\
2459377.24338348 & 15102.5 & 8.9 & NRES1 \\
2459411.24579787 & 15080.6 & 8.1 & NRES1 \\
2459578.74273868 & 14871.4 & 5.7 & NRES2 \\
2459580.74290906 & 14877.7 & 5.5 & NRES2 \\
2459584.74664902 & 14918.7 & 5.3 & NRES2 \\
2459593.79187993 & 14922.3 & 9.0 & NRES2 \\
2459597.72988968 & 14834.6 & 7.9 & NRES2 \\
2459601.75295545 & 14930.1 & 8.5 & NRES2 \\
2459602.75137792 & 14925.8 & 6.8 & NRES2 \\
2459604.79780727 & 14882.6 & 6.4 & NRES2 \\
2459609.68250136 & 14854.9 & 4.9 & NRES2 \\
\hline
2458909.692940 & 15780.0 & 27.0 & CHIRON\\
2458916.735210 & 15743.0 & 24.0 & CHIRON\\
2459428.457080 & 15786.0 & 20.0 & CHIRON\\	
\end{tabularx}

\subsubsection{NRES}
\label{sec:NRES}

We acquired TOI-784 spectra using the Las Cumbres Observatory's (LCO) Network of Robotic Echelle Spectrographs \citep[NRES;][]{Siverd2018}. LCO-NRES consists of four high-resolution optical echelle spectrographs (located in Chile, South Africa, Israel, and USA) with a resolution of $R \sim 53,000$ and a wavelength range spanning 380--860 nm. Each spectrograph is fed by two fibers: one fiber is illuminated by the stellar light coming from a 1-meter telescope at the respective site, and the second one is fed by a ThAr reference lamp. At the beginning of each night, each NRES unit automatically takes calibration images including bias, dark, flat, and ThAr frames. 

We randomly scheduled our observations with a time span of 1--4 days to avoid any observing bias \citep[][]{Burt2018} and constrained the lunar separation threshold to $30^{\circ}$ and the airmass to $<$ 1.6. Our exposure times ranged from 30 to 40 min, reaching an SNR between 40 and 90 at 5130~\AA. We collected 12 spectra in total from May to July 2021 on the NRES unit in South Africa (CPT) and 18 spectra from December 2021 to January 2022 on the unit in Chile (LSC). After binning multiple shots taken on the same nights to enhance the signal-to-noise ratio (SNR), we ended up with 20 RV data points from LCO-NRES. The raw spectra were calibrated and wavelength corrected using the CERES pipeline \citep{Brahm2017}. By cross-correlating the observed spectra with a binary mask, the CERES pipeline also delivers radial velocity measurements for each spectrum. The reported internal RV precision on TOI-784 is typically around 10~m/s. See Table~\ref{tab:rv-data-points} for all the NRES RVs.

\subsubsection{CHIRON}
\label{sec:CHIRON}

We obtained three observations of TOI-784 via the CHIRON facility between March 2020 and August 2021 to provide reconnaissance spectroscopic vetting of the target. CHIRON is a fiber-fed high-resolution echelle spectrograph at the SMARTS 1.5-m telescope located at Cerro Tololo Inter-American Observatory, Chile \citep{Tokovinin2013} with a spectral coverage of 410 to 870 nm. TOI-784 was observed in the \emph{fiber} mode, with a spectral resolving power of $R\sim$ 28,000 and an exposure time of 5 minutes. The RV precision is about 20--30~m/s.
The spectra were extracted through the official CHIRON pipeline described in \citet{Paredes2021}. We derived RVs from a least-squares deconvolution between the observation and a non-rotating synthetic template, which is generated via the ATLAS9 atmosphere models \citep{Kurucz2004} at the given spectral parameters of the targets. The derived broadening profile is fitted with a kernel accounting for the effects of rotational, macroturbulent, instrumental broadening, and radial velocity shift. The CHIRON spectra were also used to estimate stellar parameters, which is described in detail in Section~\ref{sec:spectroscopic-pars}. The CHIRON RVs are listed in Table~\ref{tab:rv-data-points}.

\subsection{High-Resolution Speckle Imaging}
\label{sec:direct-imaging}

Spatially close stellar companions can create false-positive transit signals and/or lessen the transit depth, causing the exoplanet's radius to be underestimated. Thus, determination of the ``third-light” flux contamination from the close companion stars is important to properly account for or rule out. To search for close-in bound (or line of sight) companions unresolved in \tess or other ground-based follow-up observations, we obtained high-resolution imaging speckle observations of TOI-784.

\subsubsection{SOAR HRcam}
\label{sec:soar}

\begin{figure}
\centering
\includegraphics[scale=0.55,angle=0]
{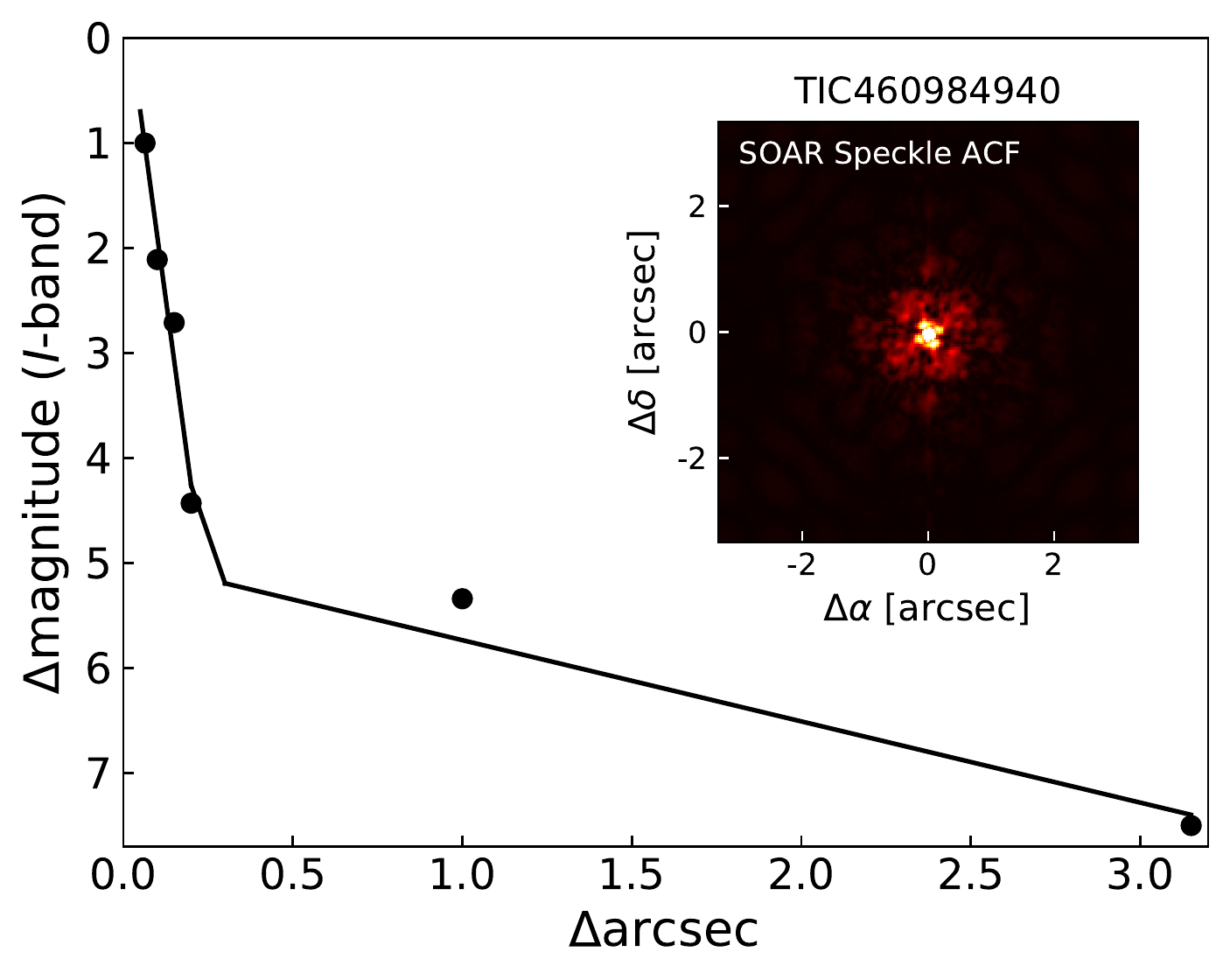}
\caption{SOAR high-resolution speckle image contrast limits in the Cousins-I band-pass. The inserts show the speckle auto-correlation function. SOAR did not detect a nearby companion to TOI-784.
\label{fig:soar-plot}}
\end{figure}

We searched for stellar companions to TOI-784 with speckle imaging on the 4.1-m Southern Astrophysical Research (SOAR) telescope \citep{Tokovinin:2018} on 14 July 2019 UT, observing in Cousins I-band, a similar visible bandpass as \tess. This observation was sensitive to a 5.4-magnitude fainter star at an angular distance of 1 arcsec from the target. More details of the observations within the SOAR \tess survey are available in \citet{Ziegler:2020}. The 5$\sigma$ detection sensitivity and speckle auto-correlation functions from the observations are shown in Figure $\ref{fig:soar-plot}$. No nearby stars were detected within 3$\arcsec$ of TOI-784 in the SOAR observations.

\subsubsection{Gemini Zorro}
\label{sec:gemini}

\begin{figure}
\centering
\includegraphics[scale=0.55,angle=0]
{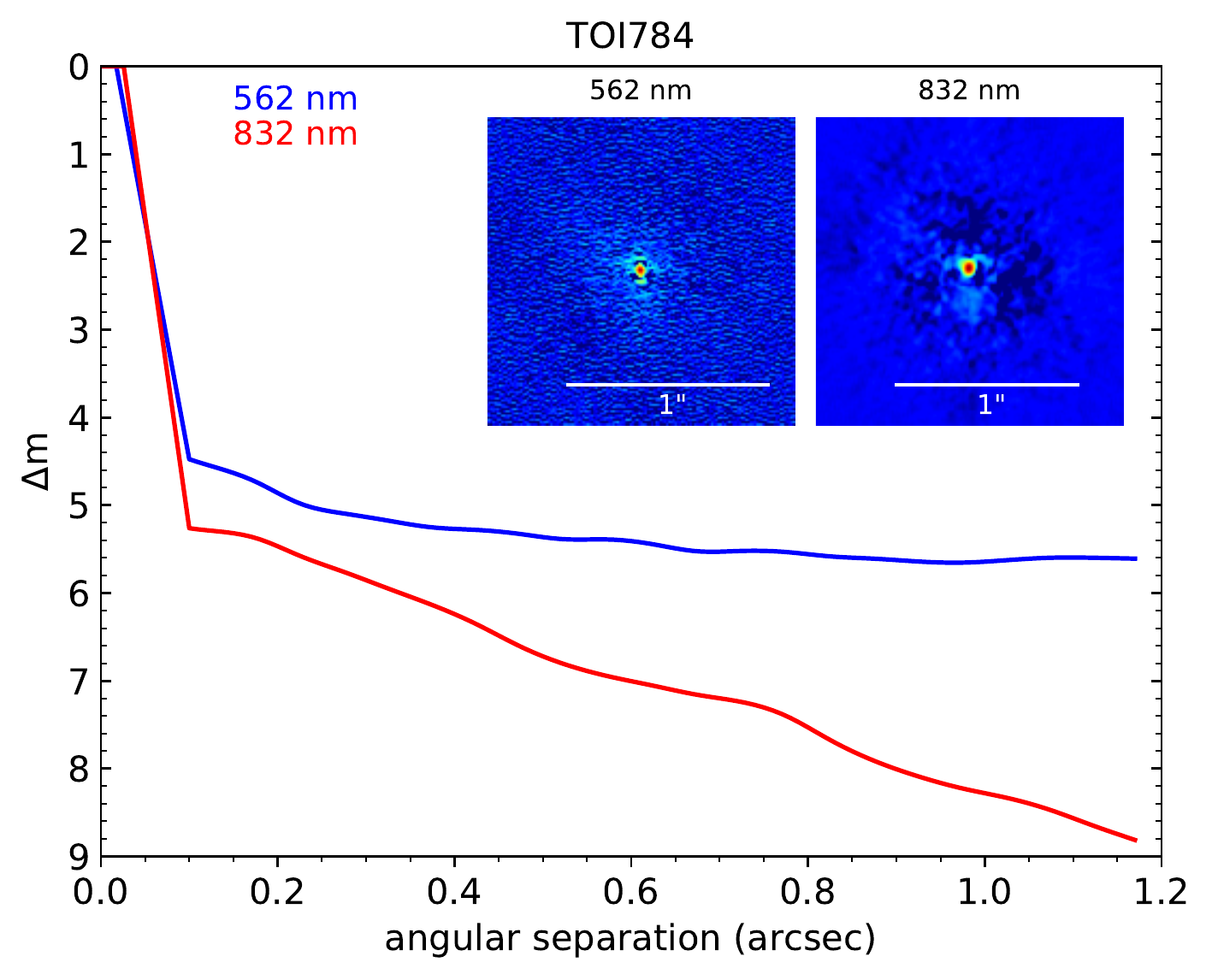}
\caption{Zorro high-resolution speckle image contrast limits in both 562 and 832 nm band-passes. The inserts show the reconstructed images in both band-passes and that TOI-784 is a single star to the contrast levels achieved (1.3 to 77.5 AU). 
\label{fig:speckle-plot}}
\end{figure}

TOI-784 was observed on 2020 March 12 UT using the Zorro speckle instrument on the Gemini South 8-m telescope\footnote {\url{https://www.gemini.edu/sciops/instruments/alopeke-zorro/}}.  Zorro provides simultaneous speckle imaging in two bands (562 nm and 832 nm) with output data products including a reconstructed image with robust contrast limits on companion detections. Three sets of $1000\times 0.06$ sec exposures were collected and subjected to Fourier analysis in our standard reduction pipeline \citep[see][]{Howell2011}. Figure~\ref{fig:speckle-plot} shows our final contrast curves and our reconstructed speckle images. We find that TOI-784 is a single star with no companion brighter than 5-9 magnitudes below that of the target star from the diffraction limit (20 mas) out to 1.2”. At the distance of TOI-784 ($d=64.6$~pc) these angular limits correspond to spatial limits of 1.3 to 77.5 AU.

\section{Stellar characterisation}
\label{sec:stellar-characterisation}
\subsection{Spectral Energy Distribution}
\label{sec:SED}

\begin{figure}
    \centering
    \includegraphics[width=0.75\linewidth,trim=75 80 80 80, clip, angle=90]{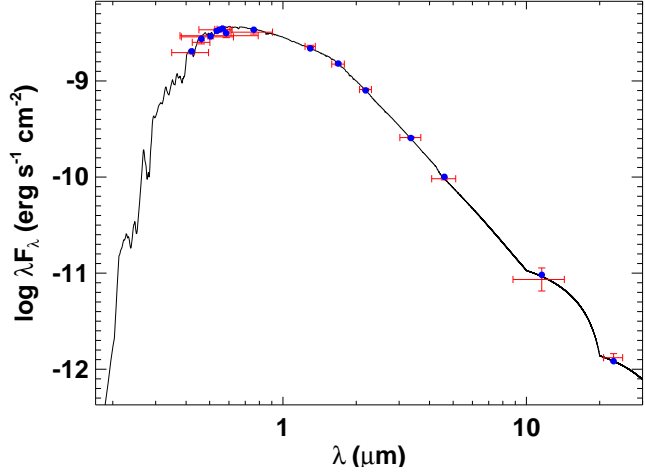}
    \caption{The best SED fit for TOI-784. Red symbols represent the observed photometric measurements, where the horizontal bars represent the effective width of the band-pass. Blue symbols are the model fluxes from the best-fit Kurucz atmosphere model (black solid line, \citealt{Kurucz1979, Kurucz2004}). 
\label{fig:sed}}
\end{figure}

As an independent check on the fundamental parameters of the host star, we carry out an analysis of the broadband Spectral Energy Distribution (SED) together with the {\it Gaia\/} EDR3 parallax in order to determine an independent, empirical measurement of the stellar radius, following the procedures described in \citet{Stassun:2016}, \citet{Stassun:2017}, and \citet{Stassun:2018}. We pull the $BV$ magnitudes from \citet{Mermilliod:2006}, the $B_T V_T$ magnitudes from {\it Tycho-2}, the $JHK_S$ magnitudes from {\it 2MASS}, the W1--W4 magnitudes from {\it WISE}, and the three Gaia magnitudes $G, G_{\rm BP}, G_{\rm RP}$. Together, the available photometry spans the full stellar SED over the wavelength range 0.4\,--\,22~$\mu$m (see Figure~\ref{fig:sed}). 

We perform a fit using Kurucz stellar atmosphere models, with the $T_{\rm eff}$, $\log g$, and [Fe/H] taken from the spectroscopic analysis (see below). The remaining parameter is the extinction ($A_V$), which we constrained to be no larger than the full line-of-sight extinction from the dust maps of \citet{Schlegel:1998}. The resulting fit is shown in Figure~\ref{fig:sed} with a reduced $\chi^2$ of 1.2 and best-fit extinction of $A_V = 0.15 \pm 0.03$. Integrating the model SED gives the bolometric flux at Earth of \mbox{$F_{\rm bol} =  5.423 \pm 0.063 \times 10^{-9}$ erg~s$^{-1}$~cm$^{-2}$}. Taking the $F_{\rm bol}$ and $T_{\rm eff}$ together with the {\it Gaia\/} parallax, with no adjustment for systematic parallax offset \citep[see, e.g.,][]{StassunTorres:2021}, gives the stellar radius as $R_\star = 0.907 \pm 0.017\ R_\odot$. We can also estimate the stellar mass empirically via $R_\star$ together with the spectroscopic $\log g$, which gives $M_\star = 0.91 \pm 0.10\ M_\odot$, and which is consistent with the value of $0.95 \pm 0.06 M_\odot$ inferred from the empirical relations of \citet{Torres:2010}. 

Finally, using $R_\star$ together with the spectroscopically-estimated rotational velocity ($v\sin i$; see below), we can infer the stellar rotation period, which gives $P_{\rm rot} / \sin i = 41.7 \pm 11.4$~d. That rotation period yields an estimated system age via empirical gyrochronology relations \citep{Mamajek2008} of $7.8 \pm 3.4$ Gyr. We caution the readers that the rotation period and age estimates are both based on the poorly constrained $v\sin i$ value from spectroscopy, and thus they serve as only a rough check to confirm that TOI-784 is an old star. We summarize the basic information of HD 307842 in Table~\ref{tab:basic-information}.

\subsection{Spectroscopic Parameters}
\label{sec:spectroscopic-pars}

To estimate the stellar parameters of TOI-784, we followed \citet{Zhou2020} and compared the CHIRON spectra against that of a library of $\sim 10,000$ observed spectra previously classified via the Spectroscopic Classification Pipeline \citep[SPC;][]{spc_Buchhave2012}. The library is interpolated via a gradient-boosting regressor implemented in the \code{scikit-learn} package. We found a best fit effective temperature of $5558\pm 100\ \rm{K}$, surface gravity of $\log g_{\star} = 4.48\pm 0.10$, and metallicity of $\rm{[Fe/H]} = -0.13\pm 0.08$ for TOI-784. The standard deviation from the stellar parameters derived from each spectrum are very small (31~K, 0.03~dex, and 0.03~dex, respectively), showing that the photon-limited uncertainties for the stellar parameters are significantly smaller than associated model-dependent uncertainties. Considering that the stellar properties are relatively Sun-like where models are well-calibrated, we quoted a minimum temperature error of 100~K, which is the floor systematic uncertainty ($\sim$ 2\% in temperature) recommended in \cite{Tayar2022}.

In addition to the spectroscopic atmospheric parameters, we also measured the star's projected rotational broadening velocity $v\sin{i}$ through a least-squares deconvolution analysis. As per Section~\ref{sec:CHIRON}, the line broadening profile is modeled via a convolution of kernels describing the rotational, macroturbulent, and instrumental broadening effects. We found that we cannot resolve the rotational broadening of the star at the given instrumental resolution, measuring a maximum broadening of $<3\,\mathrm{km\,s}^{-1}$ from the CHIRON observations. Nevertheless, we can safely conclude that the star is rotating at a rather slow speed, as we have mentioned in the GLS periodogram test of ground-based photometry that indicates a very weak rotation signal as well (see Section~\ref{sec:archival-Ground-based-Photometry}). 

Following the procedure in \cite{Lehtinen2016} (Equations (6)-(9)), we estimated the logarithmic $R'_{HK}$ of TOI-784 based on the PFS spectra. There are 13 among 35 PFS spectra having valid S-index measurements and are therefore used to calibrate the final $\log R'_{HK}$. An average value of $-4.99$ was obtained after the calculations. We then converted it to the age ($\log \tau$) of the target using Equation (3) in \cite{Mamajek2008} and find that $\log \tau \sim 9.8$. Thus it is not surprising that TOI-784 has a low measured $v\sin{i}$, given that $\log R'_{HK}$ falls into the ``inactive'' range of $-5.10$ to $-4.75$ as classified by \cite{TJHenry1996}. \rev{The S-index has a median value of $\sim$ 0.180, and it exhibits little variation with a standard deviation of $\sim$ 0.005, further indicating that TOI-784 is an inactive star.}

TOI-784 is likely to have no wide-orbiting stellar companions since we found no matches in the catalogs from \cite{Brandt2021} or \cite{Behmard2022}, which provided cross-calibrations of Hipparcos or TOIs with Gaia EDR3 to search for stellar companions.


\begin{deluxetable}{lcr}
\tablecaption{Basic information of HD 307842\label{tab:basic-information}}
\tablewidth{0pt}
\tablehead{
\colhead{Parameter} & \colhead{Value} & \colhead{Description}
}
\startdata
TIC ID$^1$ & 460984940 & \tess Input Catalog\\
TOI ID$^1$ & 784 & \tess Objects of Interest\\
R.A. (J2000)$^2$ & $10^h37^m21.88^s$ & Right Ascension \\
Dec. (J2000)$^2$ & $-63^{\circ}39\arcmin18.09\arcsec$ & Declination \\
$\mu_{\alpha}$ (mas/yr)$^2$ & $3.413\pm 0.012$ & Proper Motion\\
$\mu_{\delta}$ (mas/yr)$^2$ & $-154.788\pm 0.011$ & Proper Motion\\
$\varpi$ (mas)$^2$ & $15.4833\pm 0.0108$ & Parallax distance\\
$D$ (pc)$^2$ & $64.59\pm 0.05$ & Stellar distance\\
RV (km/s)$^3$ & $15.06\pm 0.66$ & Radial velocity\\
$T_{\rm{eff}}$ (K)$^4$ & $5558 \pm 100$ & Effective temperature\\
$\log g_{\star}$ (dex)$^4$ & $4.48 \pm 0.10$ & Surface gravity\\
$\rm{[Fe/H]}$ (dex)$^4$ & $-0.13 \pm 0.08$ & Stellar metallicity\\
$v\sin i$ (km/s)$^4$ & \rev{$\lesssim 1.1 \pm 0.3$} & Rotation speed\\
$T$$^1$ & $8.705\pm 0.017$ & T band magnitude\\
$V$$^1$ & $9.412\pm 0.003$ & V band magnitude\\
$\log R'_{HK}$$^4$ & $-4.99$ & \\
$R_{\star}$ ($\rm{R_{\odot}}$)$^4$ & $0.907\pm 0.017$ & Stellar radius\\
$M_{\star}$ ($\rm{M_{\odot}}$)$^4$ & $0.91\pm 0.10$ & Stellar mass\\
\enddata
\tablecomments{1. \cite{guerrero:TOIs2021ApJS}; 2. \cite{2020yCat.1350....0G}; 3. \cite{2018yCat.1345....0G}; 4. This work. }
\end{deluxetable}

\section{Estimates of Planetary Parameters}
\label{sec:planetary-analysis}
\subsection{Photometry analysis}
\label{sec:photometry-analysis}
\subsubsection{Transit Fit}
\label{sec:transit-fit-784b}

We refer to the planet candidate and its parameters with an index of ``c'' for convenience in the following analyses, though it is not confirmed to be a planet as our RV data do not fully cover one orbital phase.

We used the \code{Juliet} package \citep{Espinoza2019} to model the \tess PDCSAP flux after removing all NaN values and outliers. \code{Juliet} generates the transit model via \code{batman} \citep{Kreidberg2015} and supports Nested Samplers either using the MultiNest algorithm \citep{multinest2008, multinest2013, multinest2019} through either the \code{PyMultiNest} package \citep{pymultinest} or the \code{dynesty} package \citep{dynesty}. We adopted \code{PyMultiNest} in this work.

We first fitted a Gaussian Process (GP) with a Matern kernel to the data with the transit windows masked. The adjustable parameters and their priors along with the best-fit values are listed in Table~\ref{tab:priors-transit-planet-b}. \rev{The best-fit value corresponds to the median of the posterior distribution, with the lower and upper limits being defined by the 16th and 84th percentiles of the posterior distribution, respectively. The same applies throughout the rest of the paper.} We set uniform priors with widths of one day for the planet orbital period ($P_b$) and the time of conjunction ($t_{0,b}$) according to the values provided on ExoFOP, which are wide enough to not influence the results and can save the computational time compared to infinitely wide ones. The stellar density was estimated by the SED fitting and therefore received a normal prior centered at 1719.4~$\rm{kg/m^3}$ with $\sigma=212.6$. We adopted a new parameterization of $r_{1,b}$ and $r_{2,b}$ with a uniform prior between 0 to 1 to fit the planet-to-star radius ratio $p = R_b/R_{\star}$ and the impact parameter $b = (a_b/R_{\star})\cos i_b$, which follows the relations in the algorithm proposed by \cite{Espinoza2018}. This new parameterization and sampling will explore all the physically meaningful ranges for $p$ and $b$ in the ($b, p$) plane and meanwhile meet the condition of $b < 1+p$. 

\begin{figure*}
    \centering
	\includegraphics[width=170mm]{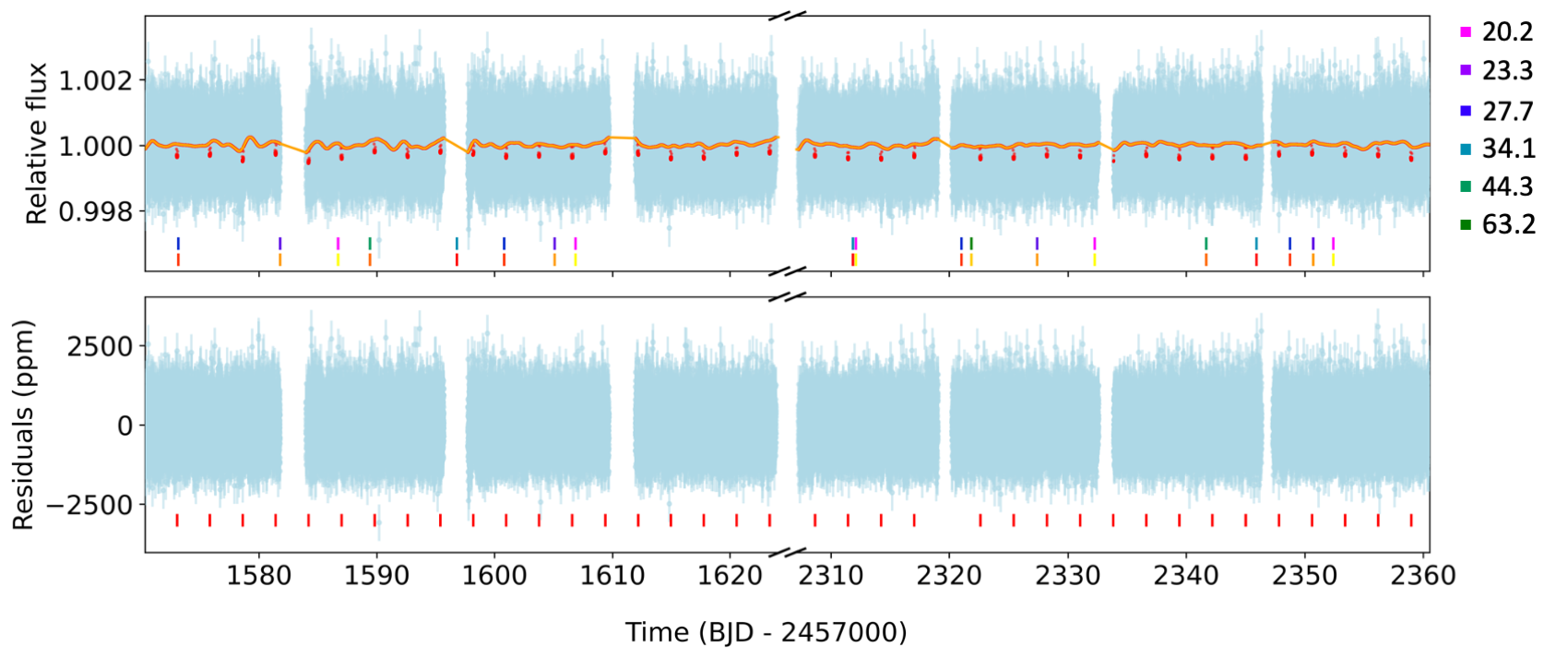}
    \caption{Top panel: \tess light curve and our best-fit detrending and transit model for planet b, with illustrations of the possible transit ephemerides of the planet candidate. The orange solid line corresponds to the detrending model using GP, while the transit fitting results are marked in red. The short vertical lines near the bottom illustrate the most probable transit ephemerides of the candidate based on the MAP grid search from \code{RadVel} (see~\ref{sec:two-planet-fit-radvel}). Each probable set of ephemerides is marked with two sets of vertical lines: the top set of magenta to cyan colors correspond to an increasing period value as labeled on the right, while the bottom set of yellow to red corresponds to an increasing MAP value. Bottom panel: Residuals of the photometric fit. In this panel only the transit events of planet b are marked with small red vertical lines. Note that the time-axis is not continuous as there is a 683-day gap with no \tess measurements. }
    \label{fig:784transit}
\end{figure*}

\begin{figure}
	\includegraphics[width=\columnwidth]{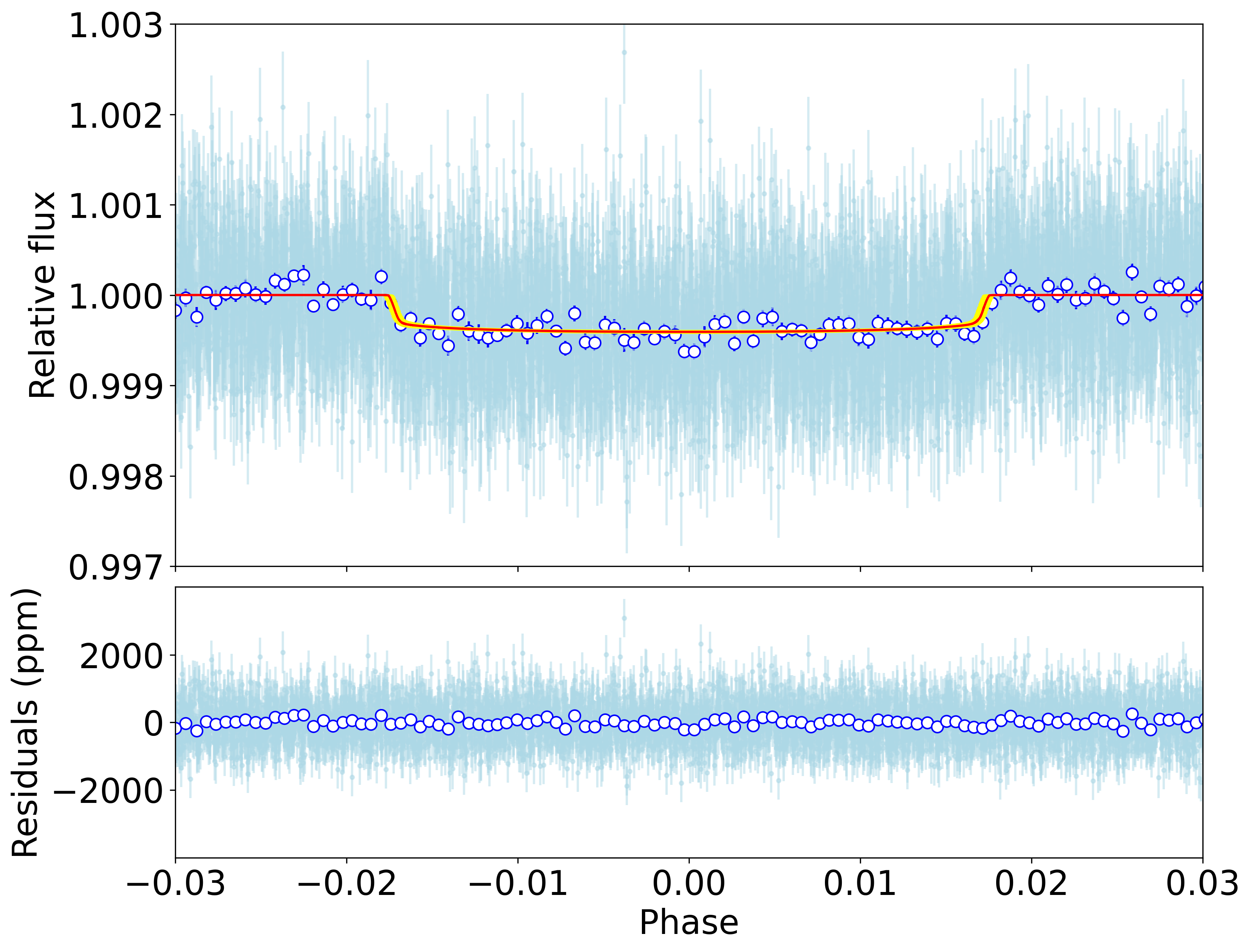}
    \caption{Phase-folded \tess light curve (light blue) and the best-fit transit model (red line with yellow 1-$\sigma$ band) for planet b, with the residuals plotted in the lower panel. The big circular points show the binned photometry with 40 data points in each bin.}
    \label{fig:784transit_phasefold}
\end{figure}

We floated the eccentricity $e_b$ and the argument of periapsis $\omega_b$ to test if the photometry could provide any constraint on the planet's eccentricity. For $e_b$ and $\omega_b$ we gave wide, uninformative priors, $\mathcal{U}(0,1)$ and $\mathcal{U}(0, 360^{\circ})$. No obvious evidence for an eccentric orbit was found and $e_b$ converged towards zero with the value of $0.10^{+0.10}_{-0.07}$. The Bayesian model log-evidence $\ln Z$ between the circular and eccentric orbit models were compared following \cite{Trotta2008}. We judge whether one model is more favored than the other by considering $2\leq\Delta \ln Z<5$ as moderate evidence and $\Delta \ln Z \geq 5$ as strong evidence. In our case, $\ln Z$ of a circular orbit model is $\sim$ 2.3 larger than the eccentric one. Given the small period of planet b, we also expect that tidal force might have damped the eccentricity to below a detectable level of \tess light curves. We thus adopted $e_b\equiv 0$. 

In principle, the PDCSAP file was already removed the flux contamination from other stars in the nearby pixels, but considering the contamination ratio estimated by TICv8 is relatively large (0.1) for this target, we floated the \tess photometric dilution factor $D_{\rm{TESS}}$ with a normal prior centered around 1.0 with a standard deviation of 0.1. This parameter converged to the value of $0.98\pm0.09$, so it was then fixed to 1.0 in the following TTV analysis (Section~\ref{sec:TTV}). We gave uniform priors $\mathcal{U}(0,1)$ to the quadratic limb darkening coefficients $q_{1,\rm{TESS}}$ and $q_{2,\rm{TESS}}$. 

Figure~\ref{fig:784transit} and~\ref{fig:784transit_phasefold} show our fitting results. We obtained a planet period of $2.797$~days for TOI-784b, which is consistent with the \tess provided value. Combining with the stellar radius in Section~\ref{sec:SED}, we derived the planet radius $R_b$ to be $1.93^{+0.11}_{-0.09}\ \rm{R_{\oplus}}$ from the best-fit values of $b$ and $p$. 

\begin{table*}
 	\centering
 	\caption{Priors, best-fit values, and derived parameters of \tess photometry analysis}
 	\label{tab:priors-transit-planet-b}
 	\begin{tabular}{llcr} 
 		\hline
 		\hline
		Parameter & Prior & Best-fit & Description\\
 		\hline
		\textit{Detrending parameters}\\
		$D_{\rm{TESS}}$ & Fixed & 1 & \tess photometric dilution factor\\
		$M_{\rm{TESS}}$ & $\mathcal{N}(0, 0.1^2)$ & $0.00010^{+0.000096}_{0.000094}$ & Mean out-of-transit flux of \tess photometry\\
		$\sigma_{\rm{TESS}}$~(ppm) & $\mathcal{J}(10^{-6}, 10^6)$ & $172.76^{+4.80}_{-4.69}$ & Extra photometric jitter term of \tess light curve\\
		$\sigma_{\rm{GP}, \rm{TESS}}$~(ppm) & $\mathcal{J}(10^{-6}, 10^6)$ & $0.00049^{+0.000029}_{-0.000015}$ & Amplitude of the Gaussian Process\\
		$\rho_{\rm{GP}, \rm{TESS}}$ & $\mathcal{J}(10^{-3}, 10^3)$ & $2.52^{+0.16}_{-0.11}$ & Length scale of the Gaussian Process\\
		\textit{Transit parameters}\\
		$P_b$~(day) & $\mathcal{U}(2.3,3.3)$ & $2.7970365^{+0.0000031}_{-0.0000030}$ & Orbital period of TOI-784b\\
		$t_{0,b}$~(BJD-2457000) & $\mathcal{U}(2336.1,2337.1)$ & $2336.61221^{+0.00044}_{-0.00050}$ & Time of transit-center for TOI-784b\\
		$r_{1,b}$ & $\mathcal{U}(0,1)$ & $0.480^{+0.081}_{-0.088}$ & Parametrisation for p and b\\
		$r_{2,b}$ & $\mathcal{U}(0,1)$ & $0.0195^{+0.0010}_{-0.0009}$ & Parametrisation for p and b\\
		$e_b$ & Fixed & 0 & Orbital eccentricity of TOI-784b\\
		$\omega_b$ & Fixed & 90 & Argument of periapsis of TOI-784b\\
		$\rho_{\star}$~($\rm{kg/m^3}$) & $\mathcal{N}(1719.4, 212.6^2)$ & $1781^{+111}_{-178}$ & Stellar density\\
		\textit{\tess photometry parameters}\\
		$q_{1, \rm{TESS}}$ & $\mathcal{U}(0,1)$ & $0.10^{+0.13}_{-0.07}$ & Quadratic limb darkening coefficient\\
		$q_{2, \rm{TESS}}$ & $\mathcal{U}(0,1)$ & $0.28^{+0.35}_{-0.21}$ & Quadratic limb darkening coefficient\\
		$D_{\rm{TESS}}$ & $\mathcal{N}(1, 0.1^2)$ & $0.981^{+0.090}_{-0.086}$ & \tess photometric dilution factor\\
		$M_{\rm{TESS}}$ & $\mathcal{N}(0, 0.1^2)$ & $-0.0000015^{+0.0000022}_{-0.0000021}$ & Mean out-of-transit flux of \tess photometry\\
		$\sigma_{\rm{TESS}}$~(ppm) & $\mathcal{J}(0.1, 1000)$ & $158.8^{+5.5}_{-5.4}$ & Additional photometric jitter term of \tess\\
        \textit{Derived planetary parameters}\\
        $a_b/R_{\star}$ & -- & $9.03^{+0.18}_{-0.31}$ & Scaled semi-major axis of the orbit for TOI-784b\\
        $b_b$ & -- & $0.22^{+0.12}_{-0.13}$ & Impact parameter of the orbit for TOI-784b\\
        $p_b$ & -- & $0.0195^{+0.0010}_{-0.0009}$ & Planet-to-star radius ratio\\
        $i_b$~($^{\circ}$) & -- & $88.60^{+0.84}_{-0.86}$ & Orbital inclination of TOI-784b\\        
 		\hline
 	\end{tabular}
\end{table*}

\subsubsection{TTV Analysis}
\label{sec:TTV}

After detrending the light curves of TOI-784, we also checked if the system has Transit-Timing Variation (TTV) signals using \code{Juliet}. We applied the \code{dynesty} package this time for higher computational efficiency. There are 20 transits in the first year (2019) and 18 transits in 2021. The ninth transit observed in 2021 was partiality captured at the beginning of Sector 38, and thus was excluded in the following analyses. The planet period $P_b$ and transit epoch $t_{0,b}$ used to calculate the ephemeris for each transit comes from the transit modeling on the \tess photometry in the previous subsection, which gives us $P_b = 2.797$~days and $t_{0,b} = 2336.6$~(BJD-2457000). We set a normal prior for each transit ephemeris that centered at the corresponding best-fitted time of conjunction with the standard deviation of 0.01. Other parameters' priors remain the same as in Section~\ref{sec:transit-fit-784b} except the \tess photometric dilution factor $D_{\rm{TESS}}$, which we simply fixed to one for efficiency concern. 

\begin{figure}
	\includegraphics[width=\columnwidth]{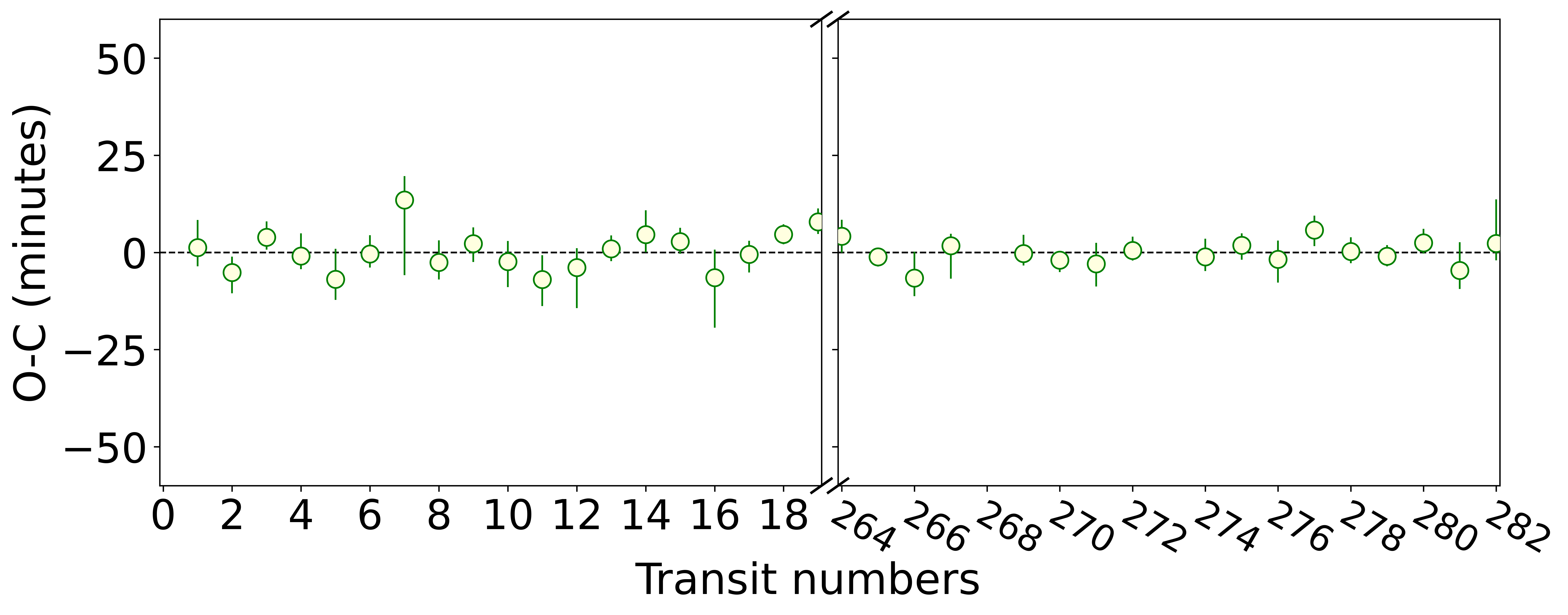}
    \caption{The $O-C$ diagram of the TTV analysis, showing the timing of the transit mid-point for each epoch. In total there were 36 transits in our TTV analysis using \code{juliet}.  We found no apparent TTV signals. }
    \label{fig:784TTV}
\end{figure}

We do not find any convincing evidence of TTV in our results. The O-Cs range from about 0.2 to 13.5 minutes with no periodic feature as shown in Figure~\ref{fig:784TTV}. This is as expected given our estimates on the possible orbital solutions and mass for the planet candidate (see Section~\ref{sec:RV}). We estimated the possible Hill radii of the candidate using the equation: $r = a_c(1-e_c)[m_c/(3M_{\star})]^{1/3}$ \citep[][]{HAMILTON199243}, and we found that the maximum Hill sphere radius of planet candidate c is $r_{H,c}\sim 0.011\ \rm{AU}$, corresponding to only $0.041(a_c-a_b)$ given $a_b\sim 0.038\ \rm{AU}$. The average ratio between $r_{H,c}$ and the separation of the two planets $(a_c-a_b)$ is about 0.035 over the six possible solutions of planet candidate c found in \code{RadVel}, which means that the second planet is too far away from planet b to cause any significant TTV.  

\subsection{RV analysis - Possible solutions of a two-planet fit}
\label{sec:RV}


\rev{Modeling the RV data taken by Magellan/PFS with a single transiting planet leaves a clear residual beyond the estimated level of stellar jitter ($\lesssim 1.2$~m/s; see Appendix~\ref{sec:one-planet-fit}). This can be seen in Figure~\ref{fig:784one_planet_fit}, which displays an example of the one-planet RV fit and shows clear residuals after subtracting the signals from planet b (in panel b). This trend is unlikely to be caused by instrumental systematics, since a significant linear trend like this, as large as $\sim$ 5~m/s in ten days, has not been observed in the RVs of any standard stars observed by PFS. For example, PFS's RVs on Tau Ceti (HD 10700) have a standard deviation of less than 2~m/s in four years. Therefore, we incorporated a second planet in our RV model.}

We performed a joint fit of the two-year PFS data along with NRES and CHIRON data using \code{RadVel}. However, due to the limited RV data timespan, we could not map out the entire orbit of the second planet. The maximum a posteriori (MAP) solution for the second planet's orbit is quite sensitive to the specific initial guesses for its RV semi-amplitude $K_c$ and orbital period $P_c$, and afterward, the results of the Markov Chain Monte Carlo (MCMC) analysis will fall into local minimums. As two out of three sections of PFS data appear to capture the same phase of the second planet's orbit (see Figure~\ref{fig:784one_planet_fit}), the degeneracy between $P_c$ and $K_c$ is not surprising. Therefore, we explore the parameter space for the potential second planet and provide a suite of best estimates for the orbital solution of this planet candidate. 


\rev{In this section, we explore the ($P_c$, $K_c$) parameter space using both \code{RadVel} and \code{Juliet} to identify potential orbital solutions for the planet candidate. Initially, we utilized the NRES and CHIRON RVs along with the PFS RVs to place an upper limit on the RV semi-major amplitude for the planet candidate, excluding the possibility of a brown dwarf. Therefore, we took four more RV observations with PFS in March 2022. With the new data at hand, we found that the possible solutions are primarily determined by the PFS RVs. To help the nested sampling converge in \code{Juliet} and simplify our model, we excluded the NRES and CHIRON RVs from subsequent analyses.} 

\begin{figure*}
    \centering
	\includegraphics[width=140mm]{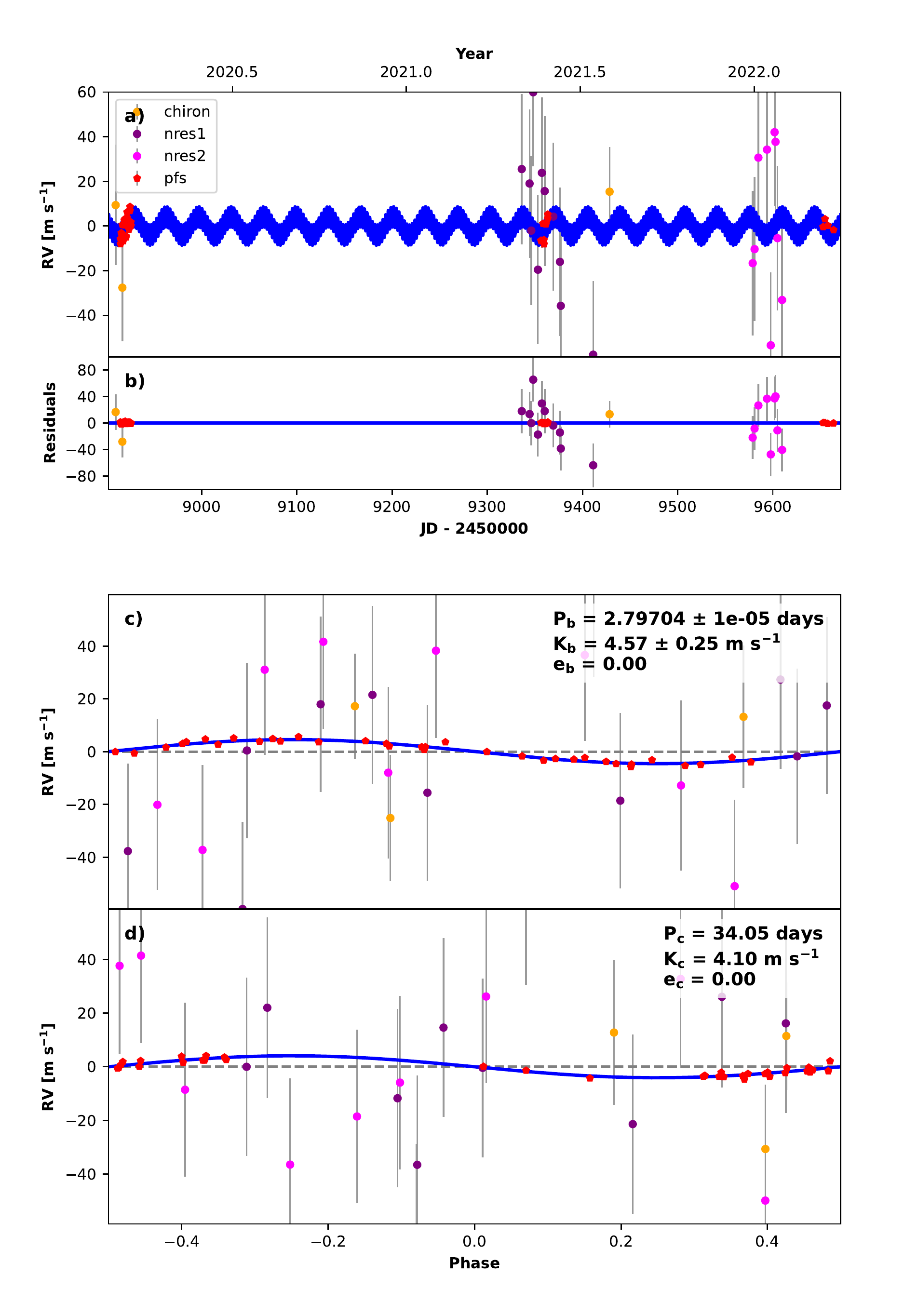}
    \caption{An example of the two-planet Keplerian orbital model. The period $P_c$ and RV semi-amplitude $K_c$ are fixed at 34.05 days, 4.1 m/s, respectively, which is the solution with the highest MAP value among our MAP grid search (see Section~\ref{sec:two-planet-fit-radvel}). RVs from different instruments are marked with different colors following previous plots: red -- PFS, purple and magenta -- NRES (indices of 1 and 2 represent facilities in South African Astronomical Observatory and Cerro Tololo Interamerican Observatory, respectively), and yellow -- CHIRON.}
    \label{fig:784_2plradvel}
\end{figure*}

\rev{Although we only use PFS data, we include an example of a two-planet fit that uses data from all three instruments in Figure~\ref{fig:784_2plradvel} for completeness. Additionally, we present our fits assuming only planet b's signal along with some other non-Keplerian signals (such as additional RV trends or a generic Gaussian Process model) in Appendix~\ref{sec:one-planet-fit}.}

\rev{Table~\ref{tab:compare-model} presents a model comparison of different radial velocity (RV) models using the Bayesian information criterion (BIC) and the second-order Akaike information criterion ($\rm AIC_c$). Both BIC and $\rm AIC_c$ are statistical model selection criteria used to assess the fit of a model to a given dataset. $\rm AIC_c$ is a corrected version of AIC that accounts for sample size. Lower values of $\rm AIC_c$ and BIC indicate a better model fit. 
According to the \code{RadVel} manual, we consider that we can not distinguish the goodness of two models when $\Delta \rm AIC_c$ is less than 2; we consider that one is slightly more favoured with $\Delta \rm AIC_c$ between 2 to 4; we consider that one is strongly disfavoured with $\Delta \rm AIC_c$ between 4 to 10; and we consider that one can be ruled out when $\Delta \rm AIC_c$ is greater than 10. 
Using these criteria, we compared the one-planet fit assuming a circular or eccentric orbit for planet b and found that the former is more favored. Thus, we applied the circular orbit and compared the one-planet fit versus the two-planet fit. We will discuss the two-planet fit in the following sections. We also incorporated a generic Gaussian process (GP) model to replace the linear trends in the one-planet fit, which is found to be strongly disfavored by the data. }

\subsubsection{Two-planet fit in \code{RadVel}}
\label{sec:two-planet-fit-radvel}

\begin{figure*}
    \centering
	\includegraphics[width=160mm]{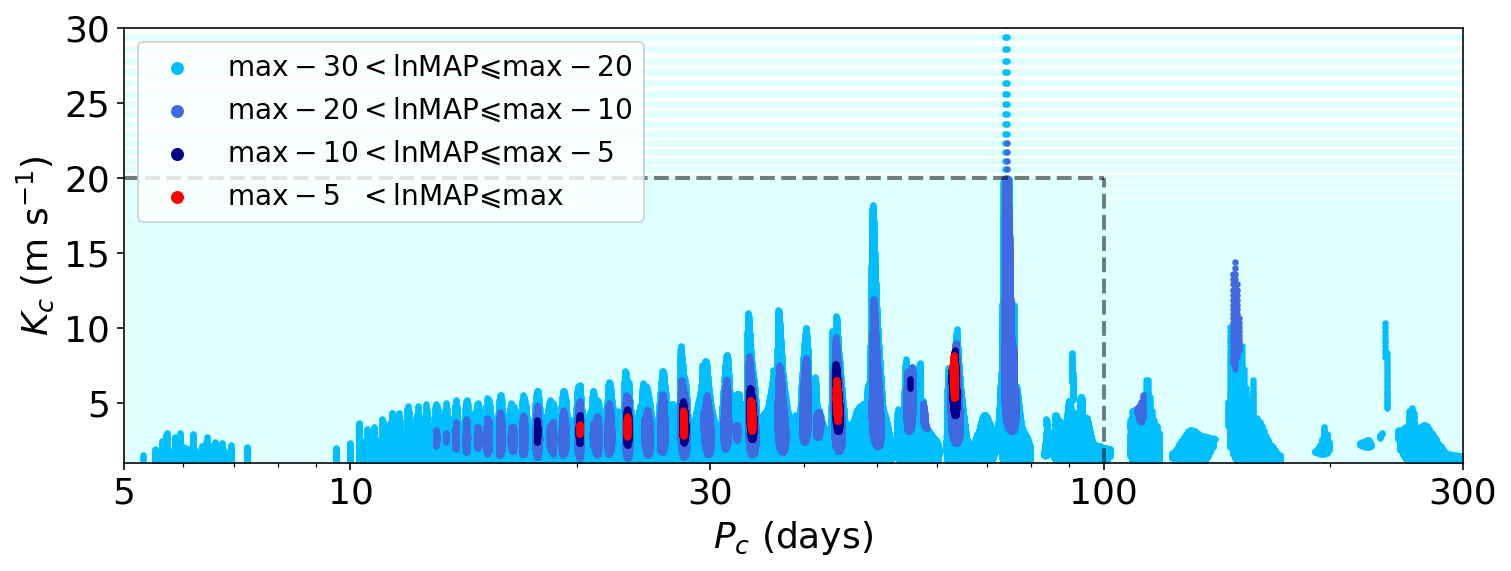}
    \caption{The $\ln$(MAP) values of a MAP grid search in the ($P_c$, $K_c$) space using \code{RadVel}. Grids placed in a log-uniform distribution were applied along $P_c$ and $K_c$ axes. The region framed by gray dashed lines was also mapped with another uniformly distributed grid with a smaller interval. As informed in the upper-left legend, different ranges of $\ln$(MAP) values are marked by different colors. We treat the periods whose $\ln$(MAP) values differ by five from the highest $\ln$(MAP) value to be ``the most possible solutions'' of the planet candidate. The cyan region is the least favored solution with $\ln$(MAP) differences larger than 30. We also explored the $K_c$ space between 30 and 60 m/s but did not find any plausible solutions, hence it is not plotted here.} 
    \label{fig:784LikelihoodMap}
\end{figure*}

\rev{We performed a grid search using the maximum a posteriori (MAP) fit method in \code{RadVel}. We constructed a grid of parameters in the ($P_c$, $K_c$) space, and for each fixed set of $P_c$ and $K_c$ values, we performed a MAP fit using \code{RadVel} with $P_b$, $t_{0,b}$, $t_{0,c}$, and the velocity zero points and jitters as the free parameters. The results are shown in Figure~\ref{fig:784LikelihoodMap}, where we color-coded the MAP values for each MAP fit using a set of ($P_c$, $K_c$) in the grid. We detail our choices for the priors and the ($P_c$, $K_c$) grid in the next two paragraphs. The most likely solutions of ($P_c$, $K_c$) are marked in red in Figure~\ref{fig:784LikelihoodMap}, corresponding to a $\ln$(MAP) difference within 5 (roughly $\Delta \rm{AIC_c}\leqslant 10$) compared to the highest one.}

\rev{The prior distributions for the free parameters are summarized in Table~\ref{tab:priors-two-planet-fit-radvel}. We gave normal distributions for $P_b$ and $t_{0,b}$ centered at the best-fit values from the photometry analysis with widths set to one order of magnitude larger than the corresponding error bar reported by the photometry. For the planet candidate, we assumed a circular orbit, and we estimated an initial guess for $t_{0,c} \sim$ BJD 2458920 by examining the residual RVs in the one-planet fit and estimating the RV zero-crossing going from redshift to blueshift (see Appendix~\ref{sec:one-planet-fit} and Figure~\ref{fig:784one_planet_fit}).}

\rev{For the ($P_c$, $K_c$) grid, we first mapped out a wider range of $(5, 1)$ (in days and m/s, respectively) to $(2000, 200)$ and found no plausible solutions beyond $\sim$ 60~m/s and 300 days. We thus narrowed down the grid to a range of $(5, 1)$ to $(300, 60)$ (days, m/s).} In order to perform a smoother and more careful search around the left corner region of the ($P_c$, $K_c$) space where a larger MAP value appears (red dots in Figure~\ref{fig:784LikelihoodMap}),
we divided the parameter space into log-uniform grids with 600 and 150 points along the $P_c$- and $K_c$-axis, respectively. We also created a more concentrated uniformly distributed grid with 0.05~m/s and 0.1~day interval on the region of $(5, 1)$ to $(100, 20)$ (gray dashed lines framed in Figure~\ref{fig:784LikelihoodMap}) to verify that varying the density of the grid will not result in additional islands of the possible solutions.\footnote{We would get more($P_c$, $K_c$) pairs falling into the red regions in Figure~\ref{fig:784LikelihoodMap}, naturally, if we choose a higher resolution for the grid, but the results should not differ in any significant way with the current grid. It is also unlikely that there would be new islands of ($P_c$, $K_c$) pairs with high MAP values, since our MAP results using \code{RadVel} are consistent with the posterior space we mapped out using \code{Juliet} in the following subsection.}.

In summary, six most likely values of $P_c$ have been revealed: $\sim$ 20, 23, 28, 34, 44, and 63 days\footnote{The same analysis was performed using joint RVs from both PFS, NRES and CHIRON for completeness, in which we found the same six best sets of $P_c$ and $K_c$ as expected.}. The MAP value of the 63-day solution is more sensitive to the chosen grid, i.e. the inputted ($P_c$, $K_c$) values compared to others, which illuminates that it should be a less stable or shallower local minimum (see also Section~\ref{sec:two-planet-fit-juliet}). We list the six ($P_c$, $K_c$) combinations with the highest MAP values in each group along with the derived $M_c\sin{i_c}$ and the corresponding $\ln$(MAP) values in Table~\ref{tab:possible-KPMR-of-c}. $K_b$ has an average of 4.7~m/s over the most possible solutions in the red islands of Figure~\ref{fig:784LikelihoodMap}, which is consistent with the one-planet fit we discussed in Appendix~\ref{sec:one-planet-fit} within the $1~\sigma$ level. This consistency demonstrates that the constraint on planet b's mass is insensitive to the orbital solution of the planet candidate. 

\begin{deluxetable*}{lcccr}
\tablecaption{Comparison of different RV fitting models$^*$ \label{tab:compare-model}}
\tablewidth{0pt}
\tablehead{
\colhead{Model} & \colhead{$\rm BIC$} & \colhead{$\rm AIC_c$} & \colhead{$\Delta \rm AIC_c$} & \colhead{Description}
}
\startdata
1pl-circular + trend & 116.07 & 111.15 & -1.23 & One-planet, circular orbit, two trends\\
1pl-eccentric + trend & 121.70 & 118.36 & 5.98 & One-planet, eccentric orbit, two trends\\
1pl-circular + GP & 159.06 & 152.27 & 39.89 & One-planet, circular orbit, GP model\\
2pl-circular & 118.71 & 112.38 & 0 & Two-planet, circular orbits for both\\
\enddata
\tablecomments{* Only PFS data were used in the fittings.}
\end{deluxetable*}

\begin{table*}
 	\centering
 	\caption{Priors for the two-planet fit MAP grid search using \code{RadVel}}
 	\label{tab:priors-two-planet-fit-radvel}
 	\begin{tabular}{llr} 
 		\hline
 		\hline
		Parameter & Prior & Description\\
 		\hline
		$P_b$~(day) & $\mathcal{N}(2.7970364, 0.00001^2)$  & Orbital period of TOI-784b\\
		$t_{0,b}$~\rev{(BJD)} & $\mathcal{N}(2459336.61220, 0.001^2)$  & Time of transit-center for TOI-784b\\
		$\sqrt{e_b}\sin\omega_b$ & Fixed to 0\\
		$\sqrt{e_b}\cos\omega_b$ & Fixed to 0\\
		$K_b\ (\rm{m/s})$ & $\mathcal{U}(-20.0, 20.0)$ & RV semi-amplitude of TOI-784b\\
		$P_c$~(day) & Fixed & Orbital period of the planet candidate\\
		$t_{0,c}$~\rev{(BJD)} & $\mathcal{U}(2457000.0, 3457000.0)$ & Time of transit-center for the planet candidate\\
		$\sqrt{e_c}\sin\omega_c$ & Fixed to 0\\
		$\sqrt{e_c}\cos\omega_c$ & Fixed to 0\\
		$K_c\ (\rm{m/s})$ & Fixed & RV semi-amplitude of the planet candidate\\
		$\dot{\gamma}$ & Fixed to 0 & Linear trend term in the RV fit\\
		$\ddot{\gamma}$ & Fixed to 0 & Curvature term in the RV fit\\
		$\mu_{\rm{PFS}}$  & $\mathcal{N}(0.0, 10.0^2)$ & Velocity zero-point for PFS\\
		$\mu_{\rm{NRES1}}$  & $\mathcal{N}(15100.0, 100.0^2)$ & Velocity zero-point for NRES1\\
		$\mu_{\rm{NRES2}}$  & $\mathcal{N}(14800.0, 100.0^2)$ & Velocity zero-point for NRES2\\
		$\mu_{\rm{CHRION}}$  & $\mathcal{N}(15700.0, 100.0^2)$ & Velocity zero-point for CHRION\\
		$\sigma_{\rm{PFS}}$ & $\mathcal{U}(0.0, 15.0)$ & Jitter term for PFS\\
		$\sigma_{\rm{NRES1}}$ & $\mathcal{U}(0.0, 60.0)$ & Jitter term for NRES1\\
		$\sigma_{\rm{NRES2}}$ & $\mathcal{U}(0.0, 60.0)$ & Jitter term for NRES2\\
		$\sigma_{\rm{CHRION}}$ & $\mathcal{U}(0.0, 60.0)$ & Jitter term for CHRION\\
 		\hline
 	\end{tabular}
\end{table*}

\subsubsection{Two-planet fit in \code{Juliet}}
\label{sec:two-planet-fit-juliet}

\begin{figure}
    \centering
	\includegraphics[width=\columnwidth]{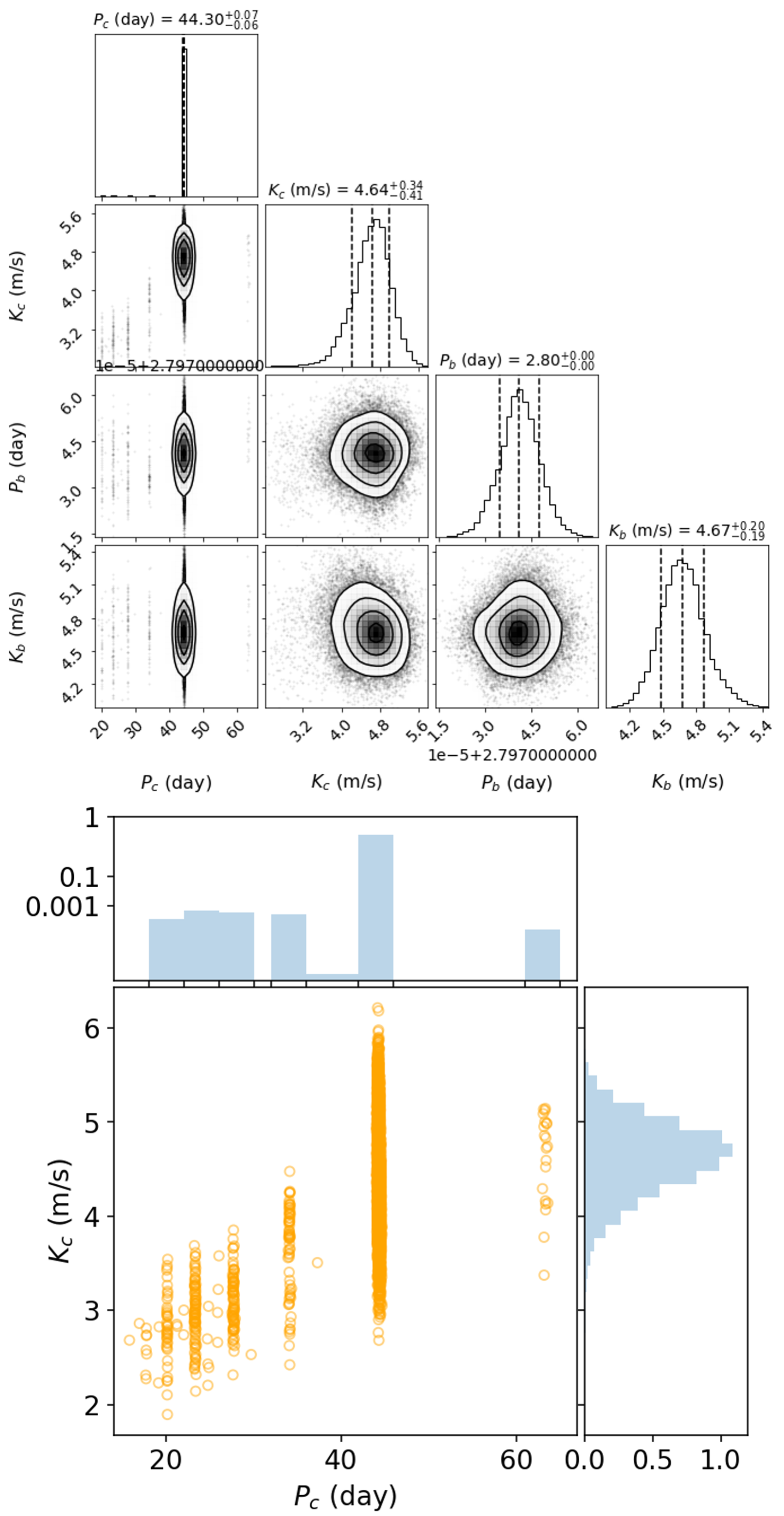}
    \caption{Top: The corner plot of the posteriors of partial variable parameters in the two-planet fit using \code{Juliet} at $n_{live}=5000$. Bottom: An enlarged version for $P_c-K_c$'s posterior space. The histogram of $P_c$ is plotted on a log scale. }
    \label{fig:784_2pl_corner_juliet}
\end{figure}

We also performed a two-planet fit using \code{Juliet} with the same priors as in our \code{RadVel} grid search, except for $K_c$ and $P_c$, where we adopted uniform priors of $\mathcal{U}(0, 60)$ and $\mathcal{U}(10.0, 450.0)$ (see Table~\ref{tab:priors-two-planet-fit-juliet}). Figure~\ref{fig:784_2pl_corner_juliet} shows the corner plot of the marginalized posterior distributions of $K_b$, $P_b$, $K_c$ and $P_c$ (a complete version with all variables included is shown in Figure~\ref{fig:784_2pl_fullcorner_juliet}). The lower panel of Figure~\ref{fig:784_2pl_corner_juliet} is an enlarged plot of $P_c-K_c$'s posterior space along with the corresponding histogram for each parameter. There are six peaks in the posterior distribution of $P_c$, which are at consistent values as the six groups we obtained in the \code{RadVel} MAP grid search (Section~\ref{sec:two-planet-fit-radvel}). 

We should note that the relative height of each possible $P_c$ in the histogram, i.e. the number of points in each island in the posterior space (orange circles in Figure~\ref{fig:784_2pl_corner_juliet}, hereafter posterior samples) do not reflect the relative probability for each solution. The fraction of posterior samples for each possible $P_c$ varies between runs even if we set the same priors and same settings for the nested sampling (including the number of live points $n_{live}$\footnote{We applied the nested sampling implemented via \code{PyMultiNest} based on the MultiNest algorithm as described in Section~\ref{sec:transit-fit-784b}. The live point stands for a series of test points uniformly distributed within the given prior for each variable parameter at the beginning, and they are then kept or rejected according to their \textit{likelihood} (see e.g. \citealt{multinest2013} for a detailed description).}). We checked the results for runs with 500, 1000, 2000, 3000, 4000, 5000, and 6000 live points and found that in general, more posterior samples were mapped out around each possible ($P_c$, $K_c$) pair with an increasing number of live points. The five islands of the ($P_c$, $K_c$) pair from 20-44 days emerge significantly more clearly in the marginalized posterior distributions after $n_{live}$ was increased to 1000, while the posterior samples around the 63-day solution only appeared when $n_{live}=5000$, which may imply that this solution is rather unstable or has a narrower or shallower local minimum in the posterior space. Since we also found the 63-day solution in the MAP grid search using \code{RadVel}, we adopted $n_{live}=5000$ and present the results in Figure~\ref{fig:784_2pl_corner_juliet}. 

\begin{figure*}
    \centering
	\includegraphics[width=125mm]{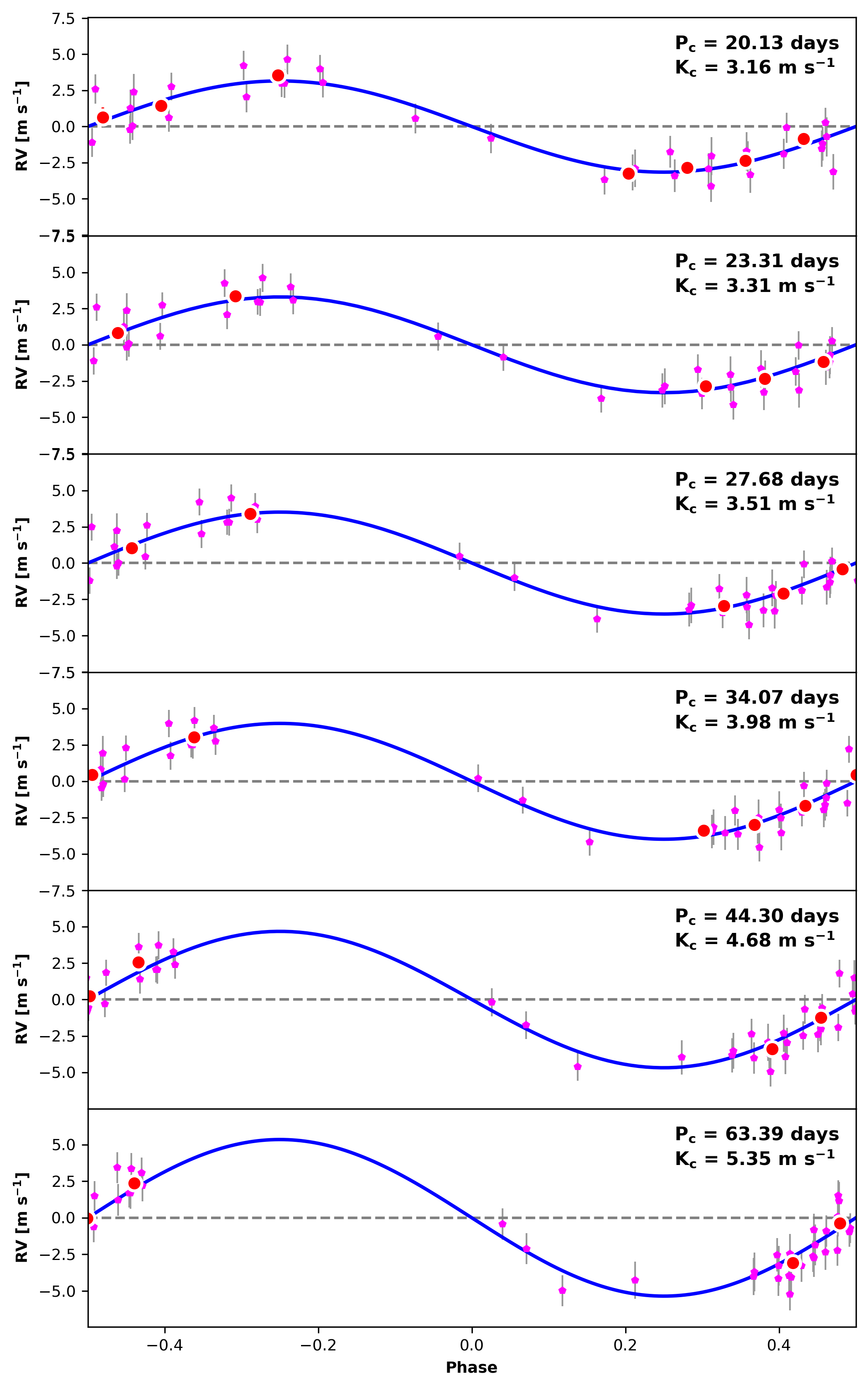}
    \caption{Phase-folded RV plots of the planet candidate with the most probable six periods from the two-planet fit in \code{Juliet}. We assumed a circular orbit for the planet candidate in all cases. PFS RVs are shown in magenta, with the big red dots representing the binned data points. }
    \label{fig:plc_phaseflod}
\end{figure*}

Though the relative probability of these six posterior peaks is unreliable \citep[e.g.,][]{Salomone2018}, the distribution within each local minimum should be trustworthy. We thus did a more concentrated search in the posterior space around each solution with \code{Juliet} to estimate the uncertainties of $P_c$ and $K_c$ of the six possible solutions. To be specific, for each of the six solutions, we set a uniform prior for $P_c$ around each solution with a width wide enough to fully cover this local minimum and away from other nearby local minimums. We also verified that varying the choice of the prior width did not affect the final posterior for each local minimum as long as the width is wide enough (e.g. one order of magnitude larger than the posterior range). The six possible solutions for the planet candidate and their estimated uncertainties from the individual posterior distributions are reported in Table~\ref{tab:possible-KPMR-of-c}. The phase-folded plots of the planet candidate are shown in Figure~\ref{fig:plc_phaseflod}, with each sub-panel showing the results of a MAP fit using ($P_c$, $K_c$) fixed at each possible solution as listed in Table~\ref{tab:possible-KPMR-of-c}. 

The orbital solution for planet b is also presented in Figure~\ref{fig:784_2pl_corner_juliet}. We derived the posterior distributions for $M_b$ and $a_b$ from the orbital parameters of planet b and the stellar parameters of TOI-784, which resulted in a median $M_b\sin{i_b}$ of $9.67\ \rm{M_{\oplus}}$ and $a_b$ of $0.04\ \rm{AU}$. These values are also consistent with what we found in the one-planet fit (Appendix~\ref{sec:one-planet-fit}) within $1~\sigma$. Considering $i_b$ derived from the transit fit which is ${88.60^{\circ}}^{+0.84}_{-0.86}$, we calculated the mass for planet b $M_b$ to be $9.67^{+0.83}_{-0.82}\ \rm{M_{\oplus}}$.

We present the properties of planet b and the planet candidate obtained from both photometry (Section~\ref{sec:photometry-analysis}) and RV analyses (Section~\ref{sec:RV}) in Table~\ref{tab:planet-pars}. 

\rev{We compare the two-planet fit with other model settings in Table~\ref{tab:compare-model}. As discussed in the preceding sections, the planet candidate has multiple orbital solutions. We report the BIC and $\rm AIC_c$ values of the model with the highest MAP value, where $P_c$ and $K_c$ were fixed at 34.05~days and 4.1~m/s, respectively (see Table~\ref{tab:possible-KPMR-of-c}). While the model of one-planet fit with the circular orbit assumption plus trends has a lower $\rm AIC_c$ than the two-planet fit, we cannot consider it the most favored model, since the combination of two linear trends is artificially included and non-physical. }

\begin{table*}
 	\centering
 	\caption{Priors and best-fit values of the two-planet fit using \code{Juliet}}
 	\label{tab:priors-two-planet-fit-juliet}
 	\begin{tabular}{llcr} 
 		\hline
 		\hline
		Parameter & Prior & Best-fit & Description\\
 		\hline
		$P_b$~(day) & $\mathcal{N}(2.7970409, 0.00001^2)$ & $2.7970343^{+0.0000063}_{-0.0000063}$ & Orbital period of TOI-784b\\
		$t_{0,b}$~\rev{(BJD)} & $\mathcal{N}(2459336.61220, 0.001^2)$ & $2459336.61221^{+0.00059}_{-0.00073}$ & Time of transit-center for TOI-784b\\
		$\sqrt{e_b}\sin\omega_b$ & Fixed & 0\\
		$\sqrt{e_b}\cos\omega_b$ & Fixed & 0\\
		$K_b\ (\rm{m/s})$ & $\mathcal{U}(-20.0, 20.0)$ & $4.67^{+0.20}_{-0.19}$ & RV semi-amplitude of TOI-784b\\
		$P_c$~(day) & $\mathcal{U}(10.0, 450.0)$ & $44.30^{+0.07}_{-0.06}$ & Orbital period of the planet candidate\\
		$t_{0,c}$~\rev{(BJD)} & $\mathcal{U}(2457000.0, 3457000.0)$ & $3079945^{+210953}_{-273077}$ & Time of transit-center for the planet candidate\\
		$\sqrt{e_c}\sin\omega_c$ & Fixed & 0\\
		$\sqrt{e_c}\cos\omega_c$ & Fixed & 0\\
		$K_c\ (\rm{m/s})$ & $\mathcal{U}(0.0, 60.0)$ & $4.64^{+0.34}_{-0.41}$ & RV semi-amplitude of the planet candidate\\
		$\mu_{\rm{PFS}}$  & $\mathcal{N}(0.0, 10.0^2)$ & $0.04^{+0.36}_{-0.40}$ & Velocity zero-point for PFS\\
		$\sigma_{\rm{PFS}}$ & $\mathcal{U}(0.0, 15.0)$ & $0.87^{+0.20}_{+0.17}$ & Jitter term for PFS\\
 		\hline
 	\end{tabular}
\end{table*}

\begin{table*}
 	\centering
 	\caption{Possible solutions of TOI-784 c found in \code{RadVel} and \code{Juliet}}
 	\label{tab:possible-KPMR-of-c}
 	\begin{tabular}{ccccc} 
 		\hline
 		\hline
 		$K_c\ (\rm{m/s})$ & $P_c\ (\rm{day})$ & $a_c\ (\rm{AU})$ & $M_c \sin{i_c}\ (\rm{M_{\oplus}})$ & $\ln$(MAP) values \\
 		\hline
        \  & \  & \textit{RadVel} & \  & \  \\
        \hline
        3.20 & 20.15 & 0.14 & 12.79 & -59.45\\ 
        3.40 & 23.30 & 0.15 & 14.26 & -56.53\\ 
        3.60 & 27.70 & 0.17 & 16.00 & -55.59\\ 
        4.10 & 34.05 & 0.20 & 19.51 & -55.52\\ 
        5.06 & 44.25 & 0.24 & 26.28 & -56.21\\ 
        7.00 & 63.20 & 0.30 & 40.95 & -57.23\\  
        \hline
        \  & \  & \textit{Juliet} & \  & \  \\
        \hline
 		$3.16^{+0.24}_{-0.24}$ & $20.128^{+0.017}_{-0.018}$ & $0.14^{+0.00}_{-0.01}$ & $12.6^{+1.4}_{-1.3}$ & --\\
 		$3.31^{+0.23}_{-0.24}$ & $23.310^{+0.023}_{-0.022}$ & $0.15^{+0.01}_{-0.01}$ & $13.8^{+1.4}_{-1.4}$ & --\\
 		$3.51^{+0.25}_{-0.26}$ & $27.682^{+0.033}_{-0.030}$ & $0.17^{+0.01}_{-0.01}$ & $15.5^{+1.6}_{-1.6}$ & --\\
 		$3.98^{+0.26}_{-0.28}$ & $34.065^{+0.045}_{-0.042}$ & $0.20^{+0.01}_{-0.01}$ & $18.9^{+2.0}_{-1.9}$ & --\\
 		$4.68^{+0.38}_{-0.40}$ & $44.296^{+0.069}_{-0.065}$ & $0.24^{+0.01}_{-0.01}$ & $24.2^{+2.8}_{-2.7}$ & --\\
        $5.35^{+0.43}_{-0.56}$ & $63.393^{+0.135}_{-0.110}$ & $0.30^{+0.01}_{-0.01}$ & $31.1^{+3.6}_{-3.9}$ & --\\
 		\hline
 	\end{tabular}
\end{table*}

\begin{table}[h]
 	\centering
 	\caption{Composite parameters for planet b and planet candidate c from photometry and RV analyses}
 	\label{tab:planet-pars}
 	\begin{tabular}{lc} 
 		\hline
 		\hline
 		Parameter & Value\\
 		\hline
 		Planet b & \\
 		$t_{0,b}$~\rev{(BJD-2457000)} & $2336.61221^{+0.00044}_{-0.00050}$ \\
 		$P_b$ (day) & $2.7970365^{+0.0000031}_{-0.0000030}$ \\
 		$K_b$ (m/s) & $4.67^{+0.20}_{-0.19}$ \\
 		$e_b$ & Fixed to 0 \\
 		$a_b$ (AU) & $0.038^{+0.001}_{-0.001}$ \\
 		$M_b$ ($\rm{M_{\oplus}}$) & $9.67^{+0.83}_{-0.82}$ \\
 		$R_b$ ($\rm{R_{\oplus}}$) & $1.93^{+0.11}_{-0.09}$ \\
 		$i_b$ ($^{\circ}$) & $88.60^{+0.84}_{-0.86}$ \\
        \rev{$\rho_b$~($\rm{g/cm^3}$)} & \rev{$7.4^{+1.4}_{-1.2}$} \\
 		\hline
 		The planet candidate & \\
 		$P_c$ (day) & 20, 23, 28, 34, 44, 63 \\
 		$K_c$ (m/s) & $3.2-5.4$ from \code{Juliet} \\
 		$e_c$ & Fixed to 0 \\
 		$a_c$ (AU) & $0.14-0.30$ \\
 		$M_c\sin{i_c}$ ($\rm{M_{\oplus}}$) & $12.6-31.1$ \\
 		$i_c$ ($^{\circ}$) & $\leqslant 88.3-89.2$ \\
 		\hline
 	\end{tabular}
\end{table}

\subsection{Does planet candidate c transit?}
\label{sec:bls}

Based on the possible $(P_c, K_c)$ provided in Section~\ref{sec:RV}, we derived the corresponding $M_c\sin{i_c}$ values and then estimated the corresponding radius, $R_c$ values, based on an empirical M-R relation with the python package \code{MRExo} \citep{lia_corrales_2015_15991}. We estimated the transit depths and durations using Equation (3) in \cite{Seager2002}. Overall, the transit depth is typically larger than 900~ppm and the duration can last for at least 4.8~hours.

Comparing the derived transit depths and durations with those of TOI-784b, whose depth and duration are roughly 400~ppm and 2.4~hours, there should be an evident signal in the \tess photometry assuming planet candidate c transits and has the best-fit parameters we found in the two-planet RV fits (Section~\ref{sec:RV}). In order to quantify the null detection of the c's transit, we performed an injection-and-recovery test based on the algorithm used in \cite{tianjun2022}. We injected a series of transit signals into the detrended \tess light curve with various sets of values for the period $P_i$ and radius $R_i$ of the planet candidate. We chose the ranges of $P_i$ and $R_i$ to cover the possible solutions we described in Section~\ref{sec:RV} with $P_i = [18, 66]$ days and $R_i = [2.8, 6.4]$ $\rm{R_{\oplus}}$. We divided the parameter space of $P_i$ and $R_i$ into a 21 by 21 grid, with each bin covering a $\Delta P_i$ of 2.4 days and $\Delta R_{i}$ of 0.18 $\rm{R_{\oplus}}$. In each bin, we generated 100 sets of random $P_i$ and $R_i$ values with random orbital phases to create the synthetic transit signals and inject them into the detrended \tess light curve of TOI-784. We then searched for any periodic signal between 15 and 70 days with a 0.01-day grid using the Box Least Squares (BLS) method \citep[][]{BLS}. This criterion is feasible because we can successfully find TOI-784b whose depth and duration are much smaller using the same criterion. We define an injected synthetic planet as ``recovered'' when the period with the highest S/N reported by the BLS search is within 0.05\% from the injected period and its harmonics (1/2 or integers) are with S/N $>10$.

\begin{figure}
    \centering
	\includegraphics[width=\columnwidth]{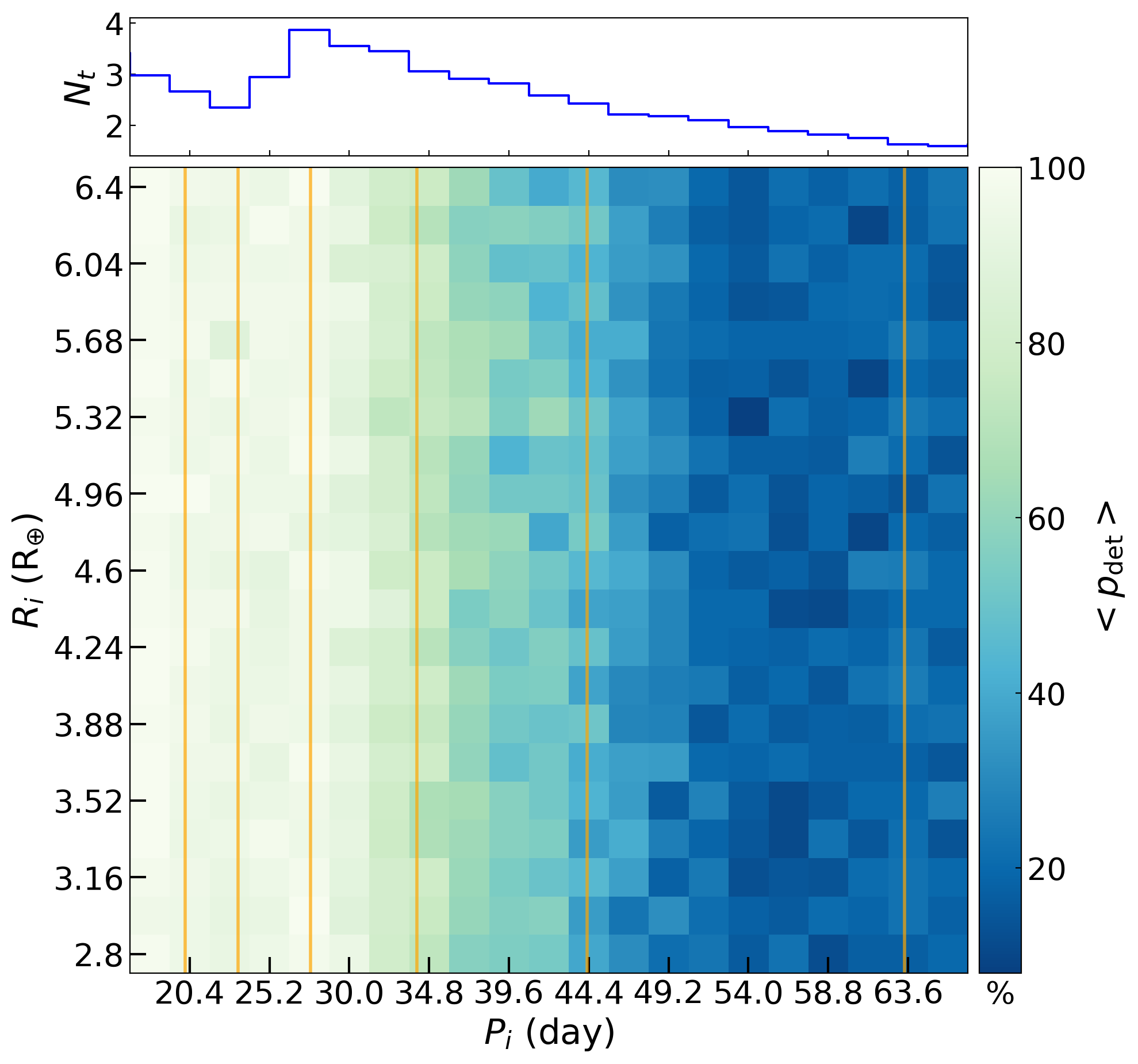}
    \caption{The result of the transit injection-and-recovery test for the planet candidate. Blue to light green represents an increasing detection possibility of its transit signal (assuming it transits). The solid blue histogram in the top panel illustrates the average number of transits within the \tess time baseline. The foreground yellow solid lines mark the six possible orbital solutions of the planet candidate found in \code{Juliet}. }
    \label{fig:784inject-and-rec}
\end{figure}

\begin{figure}
    \centering
	\includegraphics[width=\columnwidth]{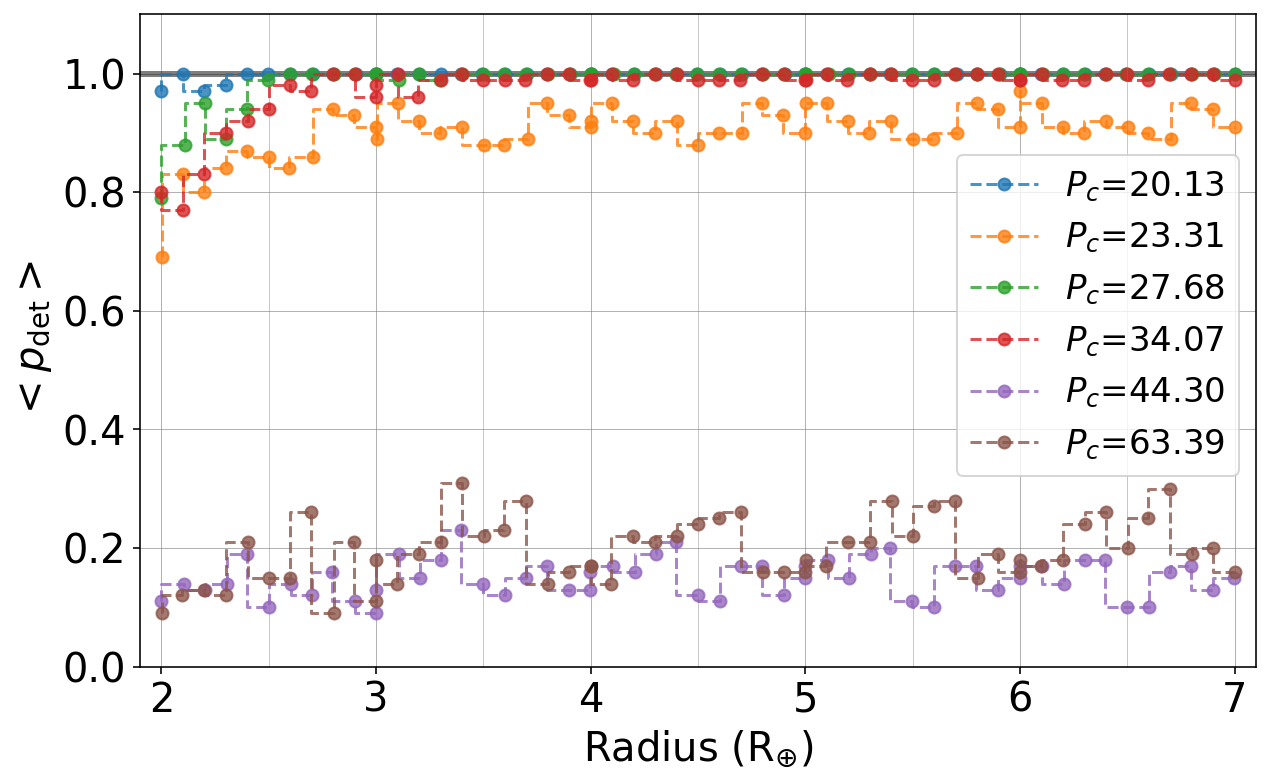}
    \caption{Detection probability as a function of the planet radius. Different colors correspond to six possible periods of the planet candidate estimated from \code{Juliet}, as shown in the legend. The results here incorporate the phase information from the RV data, and thus they are different from Figure~\ref{fig:784inject-and-rec}, which used random phase.}
    \label{fig:784inject-and-rec-2}
\end{figure}

Due to the longer period of the planet candidate and the relatively short baseline of the \tess coverage separated by a two-year gap (2 sectors spanning 55 days in 2019, and another two sectors in 2021), the detection completeness is sensitive to the total number of transits within the \tess coverage, as shown in Figure~\ref{fig:784inject-and-rec}. The typical number of transits $N_t$ for each $P_i$ bin is estimated by taking the average of the numbers of transits in the corresponding simulated transit injections. 

We also performed injection-and-recovery tests for each possible period (yellow lines in Figure~\ref{fig:784inject-and-rec}), incorporating the phase information on $t_{0,c}$ from the RV data, using the same strategy as described above. The injected periods $P_i$ were randomly picked out with a corresponding $K_i$ from the maginalized $P_c-K_c$ posterior distribution we obtained from \code{Juliet} (Section~\ref{sec:two-planet-fit-juliet}). We then estimated the corresponding time of transit-center $t_{0, i}$ for the selected ($P_c$, $K_c$) pair using the MAP fit in \code{RadVel} to include the phase information. The radius of the synthetic planet $R_i$ was given by a uniform distribution ranging from 2 to 7~$\rm R_{\oplus}$. The results of the transit detection possibility as a function of the planet size are shown in Figure~\ref{fig:784inject-and-rec-2}, which resembles the general trends in Figure~\ref{fig:784inject-and-rec}. The average detection completeness integrated over the whole radius range are 99.9\%, 90.4\%, 98.9\%, 97.8\%, 15.0\%, and 19.5\% for the six possible periods from 20 to 63~days, respectively. 

Therefore, the fact that \tess did not report a transit detection of planet candidate c suggests that it either does not transit (if having a relatively shorter period) or its transits escaped \textit{TESS}'s detection due to its long period (if $P_c > \sim 40$ days). \tess will re-observe TOI-784 in Sectors 63-64, in March-May 2023, and additional \tess data would provide more leverage to constrain the transit of this planet candidate.
 
Assuming that the planet candidate does not transit, we can estimate the possible range of its orbital inclination. Our two-planet analysis provided a group of possible values of the semi-major axis $a_c$ for the planet candidate, with the minimum and maximum values being 0.14 and 0.30~AU. These translate to an orbital inclination of $i_c \leqslant 88.3^{\circ}$ or $89.2^{\circ}$. Compared with the inclination of TOI-784b estimated from the \tess light curve, ${88.60^{\circ}}^{+0.84}_{-0.86}$, these two planets could be possibly coplanar even if the second planet does not transit.

\section{Discussion and Conclusions}
\label{sec:discussion}

\begin{figure}[h]
    \centering
	\includegraphics[width=\columnwidth]{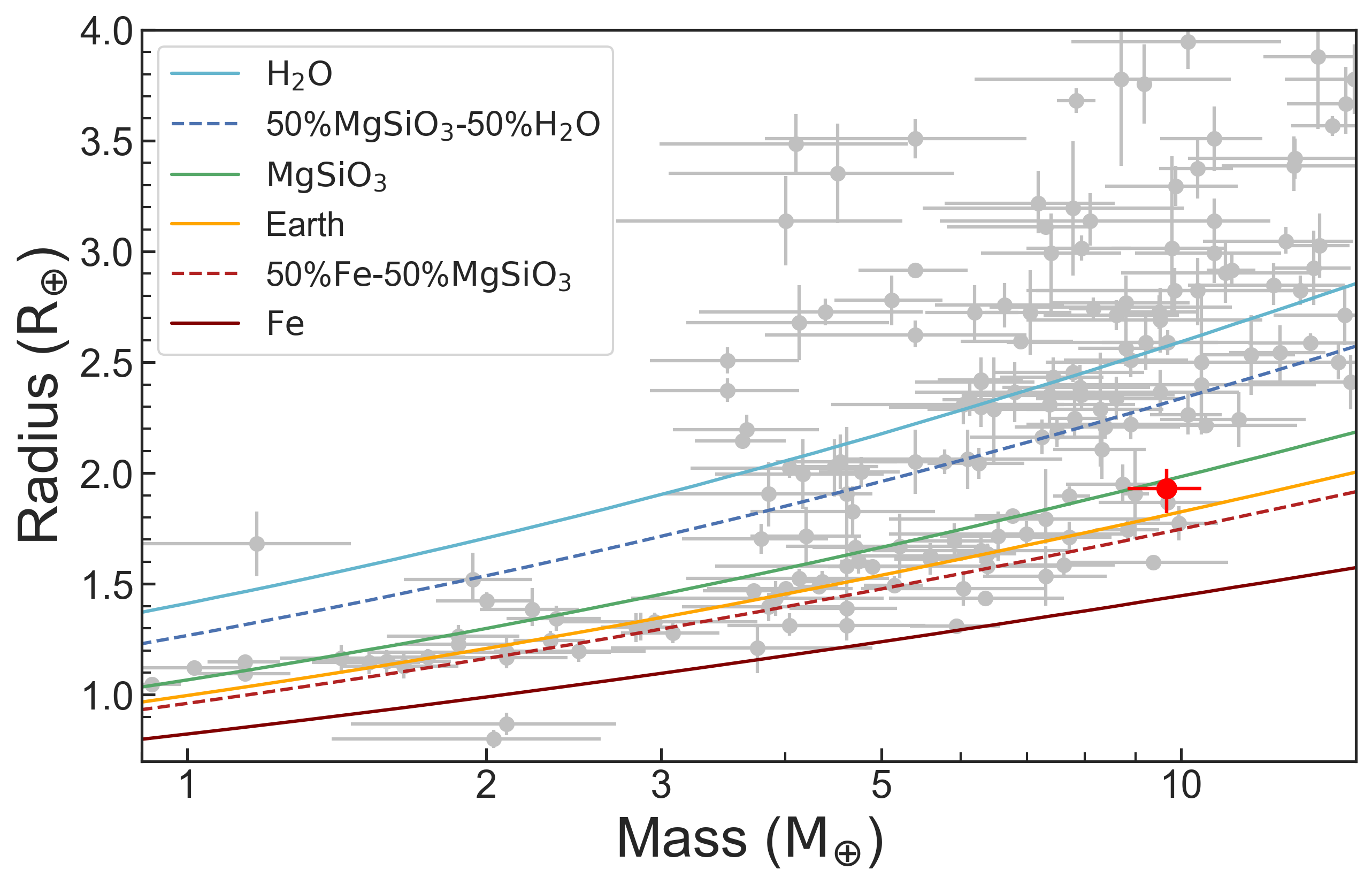}
    \caption{The M-R relation. The background gray dots are planets from TEPCAT database. Solid and dashed curves indicate different planet compositions. TOI-784b (the red point) is between the green and orange lines representing a $\rm MgSiO_3$ core and Earth composition respectively, which implies that TOI-784b is a rocky planet, the same as our analyses. Considering the uncertainty of the planet size, TOI-784b could be pure rocky or have a small amount of atmospheres consisting of heavy elements (e.g. $\rm H_2O, CO_2$, etc.).}
    \label{fig:MR_relation}
\end{figure}

Properties of planet b and planet candidate c are summarized in Table~\ref{tab:planet-pars}. In Figure~\ref{fig:MR_relation}, we show the mass-radius (M-R) diagram for the sample of confirmed exoplanets (data from TEPCAT database, \citealt{Southworth2011}) with TOI-784b marked in red.\footnote{The plot was created using the \texttt{fancy-massradius-plot} package at \url{https://github.com/oscaribv/fancy-massradius-plot} by \cite{Barragan2018}.} The solid and dashed lines illustrate theoretical models with different compositions according to \citet{zeng2016}, among which Earth composition is estimated with $\rm 34\%\ Fe + 66\%\ MgSiO_3$. The M-R relation illuminates that TOI-784b could have a rocky core and presumably no $\rm H/He$ gas envelope. The estimated density of TOI-784b using our measured radius and mass is $7.4^{+1.4}_{-1.2}\ \rm{g/cm^3}$, which also suggests that TOI-784b is a super-Earth consistent with rocky compositions and perhaps no significant atmosphere or ocean. \rev{We computed the transmission spectroscopy metric (TSM) and emission spectroscopy metric (ESM) for planet b using Equations (1) and (4) from \cite{Kempton2018}. The TSM and ESM values obtained were 36.2 and 6.8, respectively. These values fall below the threshold recommended by \cite{Kempton2018} for high-quality atmospheric characterization targets for a planet with radius of $1.5-10\ \rm R_{\oplus}$, thus may not suitable for \textit{James Webb Space Telescope} (\textit{JWST}) to observe.} However, planet b may have a small amount of volatile with heavy molecules such as $\rm H_2O$ or $\rm CO_2$ instead of $\rm H/He$ given its small size, which is not surprising given its short period and relatively strong stellar irradiation.

\begin{figure}[h]
    \centering
	\includegraphics[width=\columnwidth]{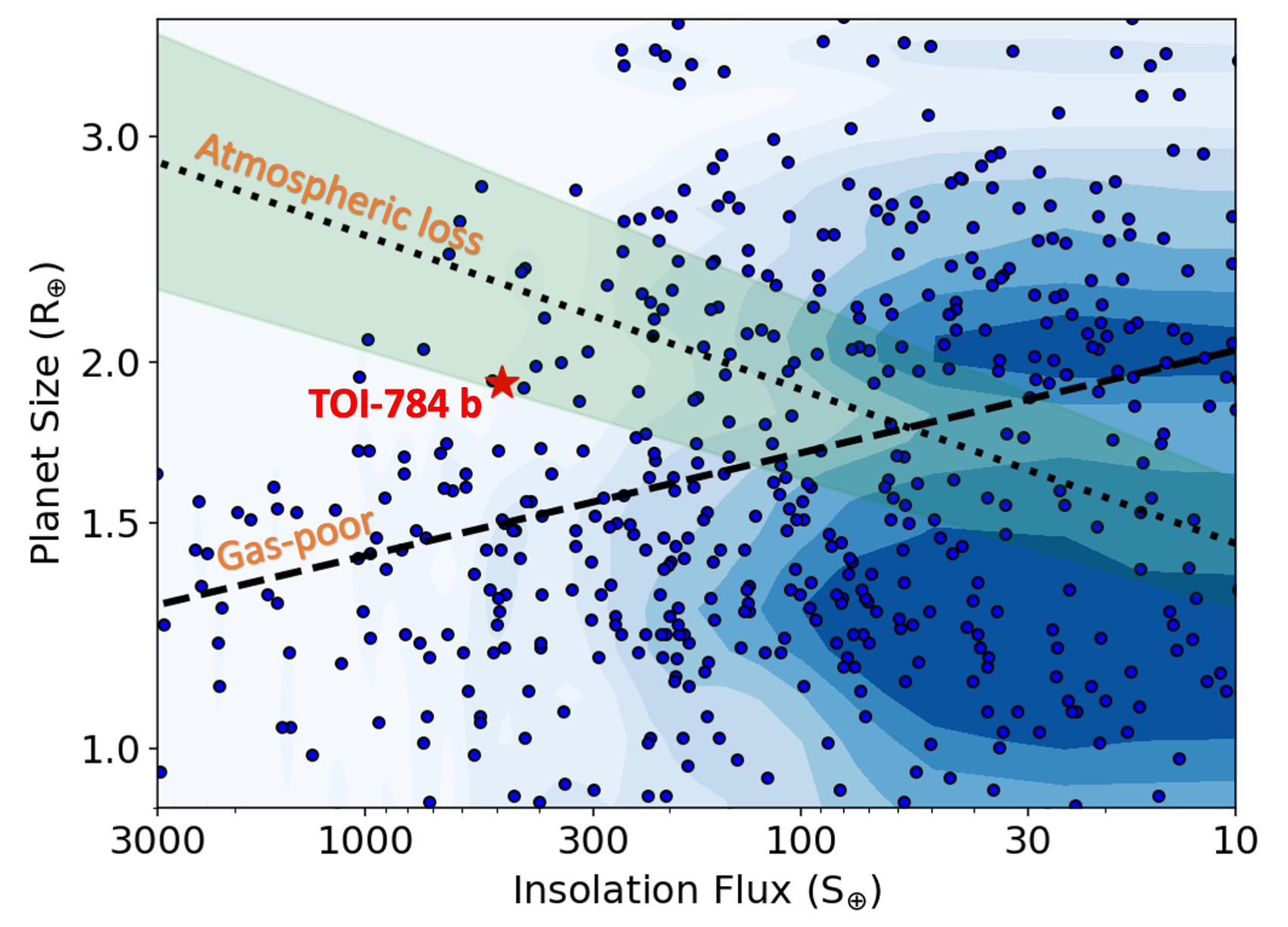}
    \caption{Relation of planet insolation flux and radius. Forward blue points are exoplanets with data taken from the NASA exoplanet archive and background contours indicate their number density. The dashed line illustrates theoretical predictions under gas-poor formation scenarios \citep{Lopez&Rice2018}, while the black dotted line is derived by \citep{Martinez2019} using the CKS sample with the shaded green region representing the $1~\sigma$ error. }
    \label{fig:insol_rade}
\end{figure}

We also compared TOI-784b with other super-Earths or sub-Neptunes using data from the NASA Exoplanet Archive\footnote{\url{https://exoplanetarchive.ipac.caltech.edu}, June 21, 2022}. Figure~\ref{fig:insol_rade} shows the insolation flux $S_{pl}$ vs.~planet radius $R_{pl}$ of TOI-784b in comparison with other known transiting planets. The background contour represents the observed planet number density (not occurrence rate). The dotted line with the $1\sigma$ errorbar shaded in light green is the empirical relation for the radius valley derived by \citet{Martinez2019}, which is consistent with both photoevaporation and core-powered mass-loss models of planet formation. The dashed black line represents the radius valley in a gas-poor formation scenario as predicted by \citet{Lopez&Rice2018}. 

TOI-784b has a radius of $1.9^{+0.11}_{-0.09}\rm{R_{\oplus}}$, which is at the upper edge of the radius valley \citep{Fulton2017, Fulton2018}. Various scenarios as mentioned above were proposed to explain this valley. A planet will receive the UV/X-ray flux from its parent star which will erode its gas envelope (photoevaporation). As a consequence, planets on a closer orbit to their hosts experience stronger irradiation and become naked rocky cores, while outer planets tend to keep their envelopes and have a transit radius roughly twice the size of the rocky core \citep{owen_wu2017}. Core-powered mass loss considered the effect of the cooling core, which can also strip envelopes and cause a lack of planets with intermediate radii \citep{Ginzburg2018, Gupta&Schlichting2020}. Photoevaporation, as well as the core-powered mass loss scenario, predicts a positive correlation in the insolation flux vs. planet size plane, while gas-poor formation makes an opposite prediction (see Table 5 and Figure 11 in \citealt{Cloutier&Menou2020} for a summary) as shown in Figure~\ref{fig:insol_rade}. TOI-784b is located at the lower edge of the radius valley (in the $R_p - S_p$ plane) predicted by photoevaporation or core-powered mass loss, which is consistent with the atmospheric loss formation scenario considering its rocky composition. 

Moreover, the well-separated planet candidate and TOI-784b (period ratio $>7$) may result from a violent mass loss history \citep{SuWang2023}. According to previous work on the stability of multi-planet systems with mass loss process, it is possible to keep the system in a stable configuration if the inner planet loses less than 5-10\% of its total mass in a timescale larger than $2 \times 10^4$~years \citep{Matsumoto2020, SuWang2023}. After such ``gentle'' mass loss process, a closely packed system could still remain stable. On the other hand, if the inner super-Earth experiences a violent atmosphere mass loss in a shorter timescale, in which about 20-30\% of its total mass is lost, a widely separated configuration may form as a consequence of violent dynamical interactions, including collisions and merges between planets. Therefore, the well-separated configuration of TOI-784 system is consistent with the possibility that the inner planet once held a considerably amount of atmosphere.

In addition to the mass loss formation scenario for TOI-784b, there is another possibility that this rocky planet could be a failed core and preserved as a super-Earth in the natal disk \citep[e.g.,][]{Ida2004}. According to \cite{Chen2020}, even though a massive core reaches the pebble-isolation mass (few to $10 M_{\oplus}$) before the severe depletion of the gas disk (which is different from formation in a gas-poor environment), an increasing grain opacity in the planetary envelope could significantly suppress the runaway gas accretion given its stellar proximity.

In summary, we confirmed a \tess discovered transiting super-Earth, TOI-784b, and identified an additional planet candidate in the system using radial velocities taken by Magellan/PFS, CHIRON, and LCOGT/NRES. TOI-784b is a $1.93\ \rm{R_{\oplus}}$ super-Earth, with a period of 2.80 days and mass of $9.67\ \rm{M_{\oplus}}$. The bulk density of TOI-784b indicates that its composition is likely rocky, with a minimum amount of volatiles, which is consistent with the prediction of the photoevaporation or core-powered mass loss scenario. We also found a planet candidate revealed by the RV follow-up observations, but unfortunately, our observed RVs did not map out its entire orbit. We performed a grid search in parameter space using \code{RadVel} as well as a two-planet fit  using \code{Juliet} to investigate possible solutions for the planet candidate (Section~\ref{sec:RV}). Six possible orbital solutions for the candidate have been found, with the period ranging from 20 to 63 days and mass being $12-31\ \rm{M_{\oplus}}$. Given its large minimum mass, the candidate is likely a Neptune-mass planet rich in volatiles, unlike TOI-784b. We performed an injection-recovery test to characterize the likelihood of the planet candidate actually transiting but being missed by \textit{TESS}, and we conclude that it is more likely to be an intrinsically non-transiting planet, although we cannot rule out the transit scenario completely, especially if it has a period around or longer than 40 days.

The TOI-784 system contains a transiting small planet in the inner orbit and an outer, likely warm, planet candidate that probably does not transit, which can contribute to the statistical sample of multi-planet systems with mass measurements. The determination of the outer planet candidate's properties requires future observations, especially more \tess observations in the Extended Mission and more RV follow-up.  

\section{Acknowledgments}
We thank Richard Grumitt for his insightful advice on MCMC and Nested Sampling for highly multi-modal scenarios. 
We thank David Latham for his helpful suggestions about RV modeling. 
We thank Marshall Johnson for his contributions to NRES RVs reduction.
We thank Xiaochen Zheng for her helpful discussions about the planet formation theory.
D. D. acknowledges support from the TESS Guest Investigator Program grants 80NSSC21K0108 and 80NSSC22K0185.
Hua and Wang acknowledge support from NSFC grant 12273016.
We acknowledge the use of public TESS data from pipelines at the TESS Science Office and at the TESS Science Processing Operations Center.
Resources supporting this work were provided by the NASA High-End Computing (HEC) Program through the NASA Advanced Supercomputing (NAS) Division at Ames Research Center for the production of the SPOC data products.
This paper includes data collected with the TESS mission, obtained from the MAST data archive at the Space Telescope Science Institute (STScI). Funding for the TESS mission is provided by the NASA Explorer Program. STScI is operated by the Association of Universities for Research in Astronomy, Inc., under NASA contract NAS 5–26555.
All the \tess data used in this paper can be found in MAST: \dataset[https://doi.org/10.17909/fwdt-2x66]{https://doi.org/10.17909/fwdt-2x66}. 
This research has made use of the Exoplanet Follow-up Observation Program website, which is operated by the California Institute of Technology, under contract with the National Aeronautics and Space Administration under the Exoplanet Exploration Program (doi: 10.26134/ExoFOP3).
This research has made use of the NASA Exoplanet Archive, which is operated by the California Institute of Technology, under contract with the National Aeronautics and Space Administration under the Exoplanet Exploration Program.
This research uses data obtained through the Telescope Access Program (TAP), which has been funded by the TAP member institutes.
This work makes use of observations from the Las Cumbres Observatory global telescope network.
We acknowledge LCO Key Project (KEY2020B-005) which has kindly offered observing time for the target.  
This research has used data from the CTIO/SMARTS 1.5m telescope, which is operated as part of the SMARTS Consortium by RECONS (\url{www.recons.org}) members Todd Henry, Hodari James, Wei-Chun Jao, and Leonardo Paredes. At the telescope, observations were carried out by Roberto Aviles and Rodrigo Hinojosa.

\vspace{5mm}
\facilities{\textit{TESS}, Magellan(6.5m)-PFS, LCOGT(1m)-NRES, CTIO(1.5m)-CHIRON}


\software{astropy \citep{astropy2013, astropy2018, astropy2022}, emcee \citep{emcee}, AstroImageJ \citep{Collins2017}, lightkurve \citep{Lightkurve}, RadVel \citep{Fulton2018_Radvel}, Juliet \citep{Espinoza2019}, MRExo \citep{lia_corrales_2015_15991}
          }

\rev{Dataset: All the \tess light curves used in this paper can be found in MAST: \dataset[10.17909/2rfj-4m38]{http://dx.doi.org/10.17909/2rfj-4m38}. All data we used from ExoFop can be found in \dataset[10.26134/ExoFOP3]{https://doi.org/10.26134/exofop3}. All data we used from NASA Exoplanet Archive can be found in \dataset[10.26133/NEA13]{https://doi.org/10.26133/nea13}.}




\bibliography{2021reference}{}
\bibliographystyle{aasjournal}
\clearpage

\appendix
\section{One-planet fit}
\label{sec:one-planet-fit}

Due to a lack of RV data, we cannot fully constrain the orbit of the additional planet candidate (which we called ``planet c" for simplicity in the paper), planet b's solution would be dependent on the choice of the specific two-planet fit. In this appendix, we explore the robustness of planet b's parameters and check if we could constrain the mass of planet b regardless of the choice for the orbital solution of the planet candidate.

Before we obtained the 4 more data points in 2022 from PFS, the PFS data  had unfortunately only captured the same rising phase of the second planet separated by a year, which makes it challenging for the RV analysis due to degeneracy in the period and RV amplitude parameter space for the second object. Therefore, we first constructed a composite model with a linear trend for each of the first two sections of PFS data plus a Keplerian model for planet b, which would be a more generic model describing the signals from the second object more commonly used in the RV fit for systems with long-period massive companions (e.g., \citealt{crepp2012,montet2014}). Here we present the result of our fit to the pre-2022 PFS RVs using this simple model that consists of one Keplerian model plus two trends. An alternative model including both a linear trend and a curvature has also been tested, but model comparison with BIC and $\rm AIC_c$ indicates that this was not the favored model. 

\begin{figure*}
    \centering
	\includegraphics[width=140mm]{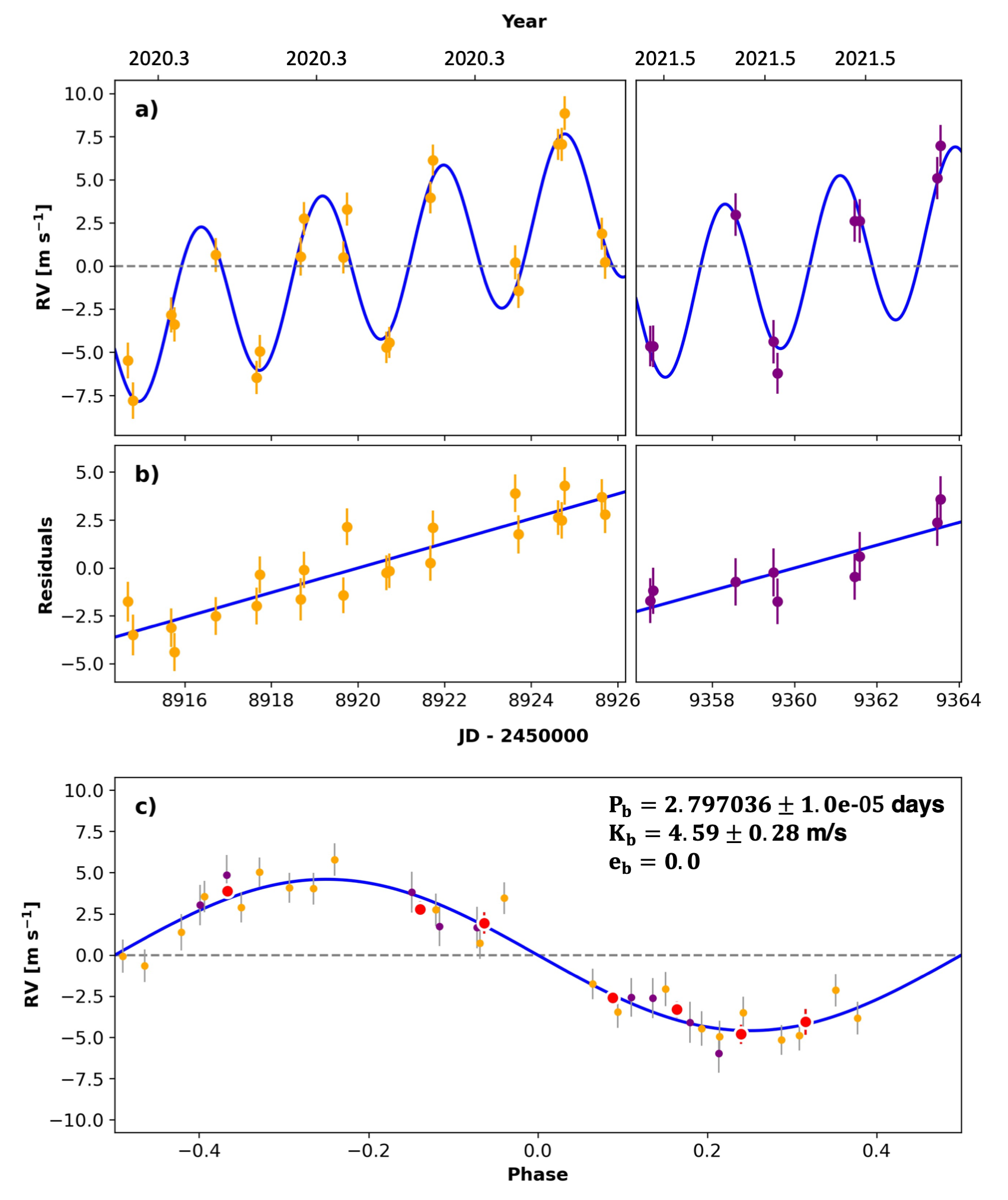}
    \caption{The one-planet RV fit with two independent trends in the model. The orange and purple dots represent the first and second year PFS data, respectively. The best-fit model is shown by the thin blue line. (b) The linear residuals after subtracting signals from TOI-784b. (c) The phase-folded RV plot of TOI-784b. Red circles are the same velocities binned in 0.08 units of the orbital phase. }
    \label{fig:784one_planet_fit}
\end{figure*}

The best-fit Keplerian $+$ 2 trends model is shown in Figure~\ref{fig:784one_planet_fit}. Similar to Section~\ref{sec:two-planet-fit-radvel}, we used \code{RadVel} \citep{Fulton2018_Radvel} to perform this fit, where a \rev{maximum a posteriori optimization} (MAP) will first be run in a fit via \code{scipy.optimize.minimize}, followed by a MCMC analysis via emcee \citep{emcee} to estimate the uncertainty for each parameter. In our case, we required the model to fit two independent linear trends, which is not supported by \code{RadVel}, so we modified its source code and added an additional term in the model.

\begin{figure*}
    \centering
	\includegraphics[width=160mm]{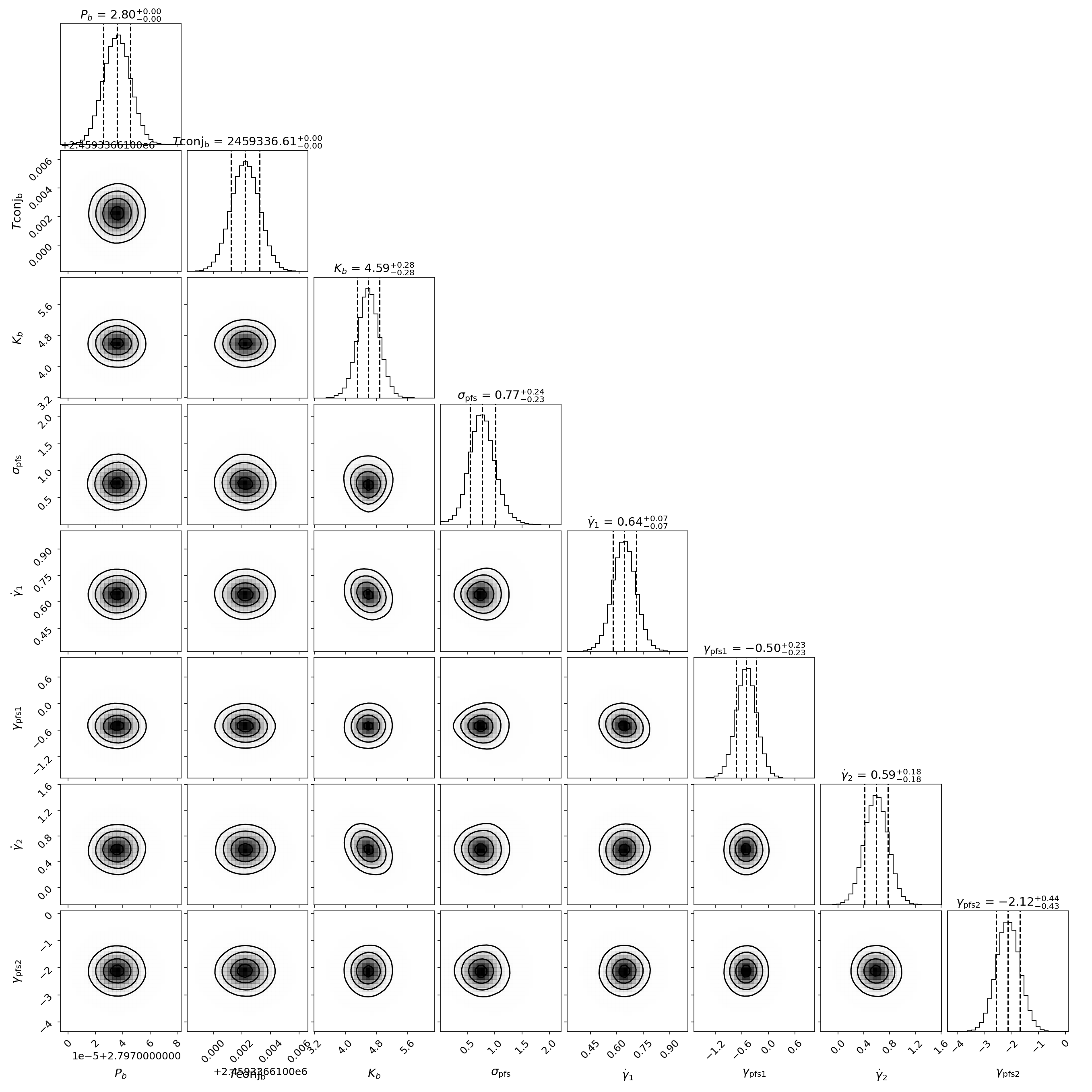}
    \caption{Corner plot of all fitted parameters in one-planet RV fit using \code{RadVel}. }
    \label{fig:784_1pl_fullcorner_radvel}
\end{figure*}

\begin{figure}
    \centering
	\includegraphics[width=80mm]{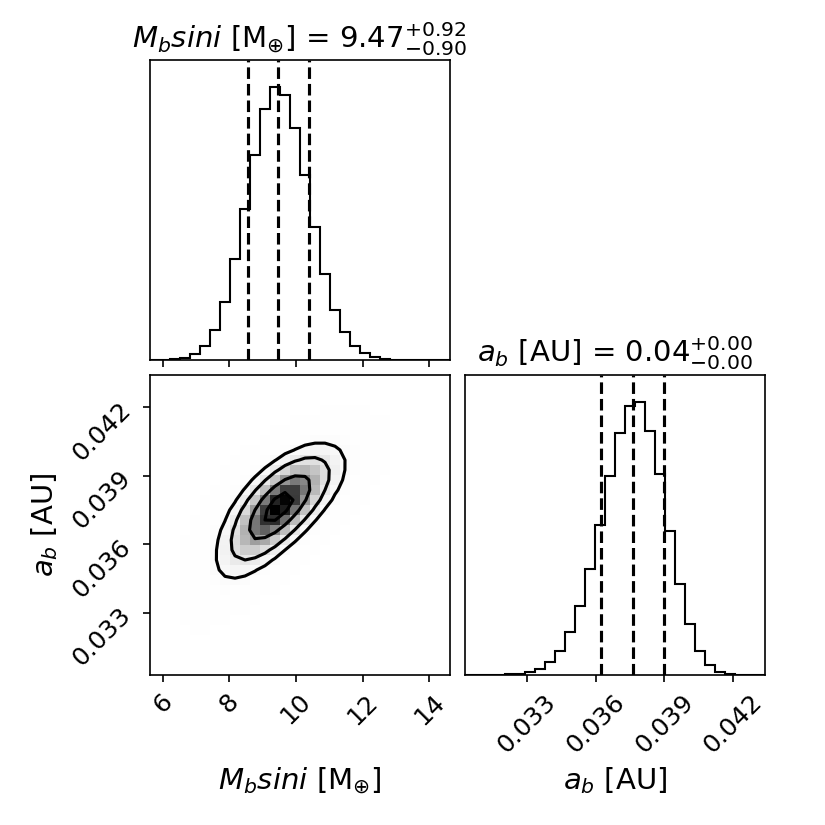}
    \caption{Corner plot of derived parameters in one-planet RV fit using \code{RadVel}. }
    \label{fig:784_1pl_derivedcorner_radvel}
\end{figure}

Priors and best-fit values are listed in table~\ref{tab:priors-one-planet-fit}. To be specific, we gave normal distributions for the orbital period $P_b$ and the time-of-transit center $t_{0,b}$ according to the optimized values we obtained in the previous photometry analysis (Section~\ref{sec:photometry-analysis}). Eccentricity was fixed to zero as we discussed in Section~\ref{sec:transit-fit-784b}. The initial guess of $K_b$ was 5~m/s and we set a uniform distribution of $\mathcal{U}(-20, 20)$ for its prior, which is roughly twice the RV range of PFS data used here. We gave two different slopes $\dot{\gamma}_1$, $\dot{\gamma}_2$ for the two sections of RV data with initial guesses both set to zero. The two sections of data share a same jitter term $\sigma_{\rm{PFS}}$ but having independent offset terms $\mu_{\rm{PFS},1}$, $\mu_{\rm{PFS},2}$ of PFS, for which we gave normal distribution $\mathcal{N}(0, 10^2)$ and two uniform distributions $\mathcal{U}(0, 15)$, respectively. The best-fit RV semi-amplitude of planet b $K_b$ and its uncertainties are $4.59^{+0.28}_{-0.29}$~m/s, and the corresponding minimum mass is $9.47^{+0.92}_{-0.90}\ \rm{M_{\oplus}}$, consistent with the values we presented in the main text of this paper (Table~\ref{tab:planet-pars}). Corner plots for the marginalized posteriors of the fitted and the derived parameters are presented in Figure~\ref{fig:784_1pl_fullcorner_radvel} and Figure~\ref{fig:784_1pl_derivedcorner_radvel}. 

\begin{figure*}[h]
    \centering
	\includegraphics[width=160mm]{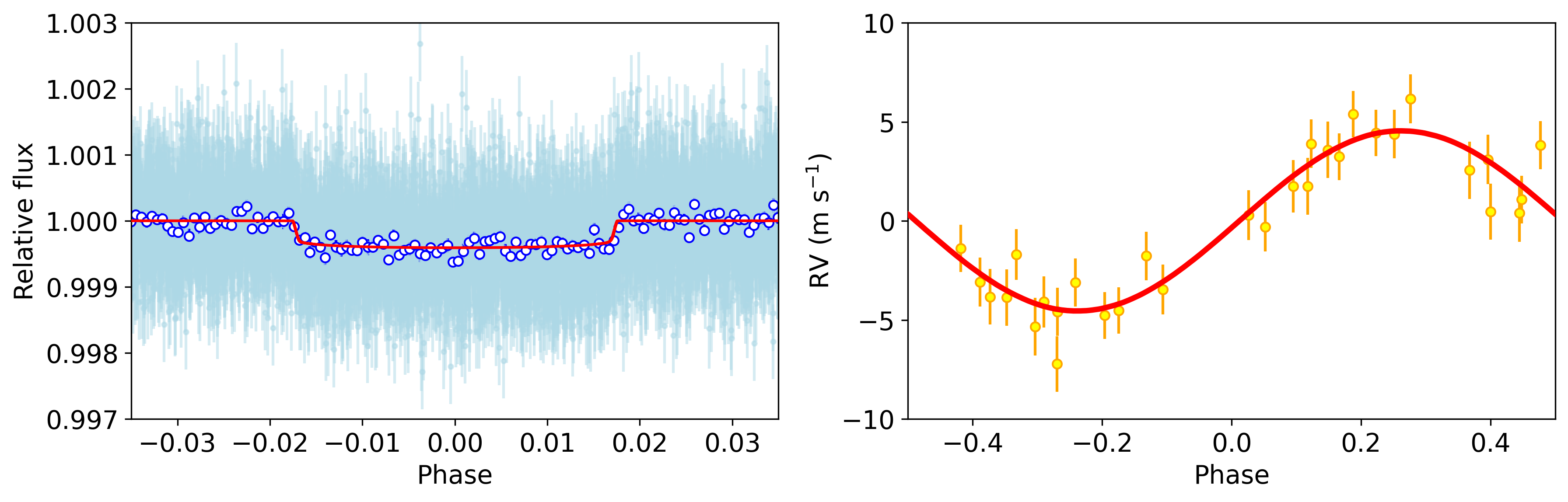}
    \caption{The joint fit of the \tess light curve and RVs of planet b using pre-2022 PFS RVs with the two best-fit linear trends subtracted. The transit fit is in the left panel, and the RV fit is on the right, with the best-fit model in red solid lines. The white dots circled by blue rings are binned \tess photometric data points with a 40 bin size. }
    \label{fig:784joint_fit}
\end{figure*}

We then subtracted the best-fit linear trends from the RV data and performed a joint fit using \code{Juliet} \citep{Espinoza2019} combining the residual RV data with the \tess light curve (Figure~\ref{fig:784joint_fit}). The joint fit did not tighten the constraints on any of the parameters for planet b, and the results are consistent well within their error bars with our results presented above (Section~\ref{sec:planetary-analysis}).

\begin{figure*}
    \centering
	\includegraphics[width=140mm]{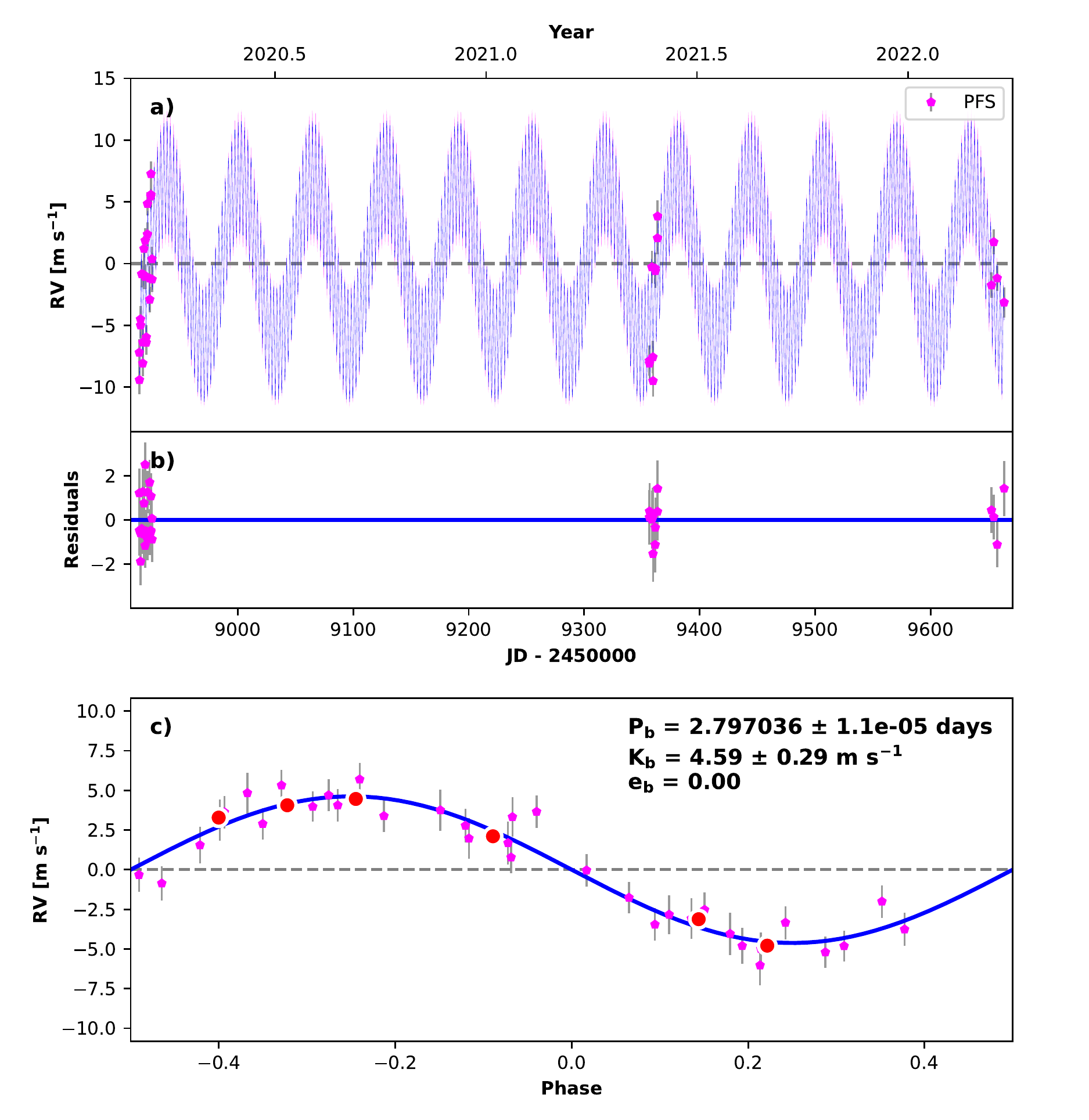}
    \caption{The one-planet RV fit with a GP model using \code{RadVel} and \code{celerite}. The magenta points are the used PFS RVs. The best-fit model is shown by the thin blue line. (b) The RV residuals after subtracting the best fit GP and Keplerian models. (c) The phase-folded RV plot of TOI-784b after subtracting the GP model component. Red circles are the same velocities binned in 0.08 units of orbital phase. }
    \label{fig:784one_planet_fit_GP}
\end{figure*}

We also performed a one-planet fit in \code{RadVel} including a Gaussian Process (GP) model to fit the additional RV signals besides TOI-784b. A GP model is a commonly used generic non-parametric model describing periodic or quasiperiod signals, more often used to describe RV jitter caused by stellar magnetic activities (e.g., \citealt{Haywood2014}). Here we chose to use the GP regression to model the RVs from the planet candidate as it is a flexible and generic time-series model that does not bear any astrophysical meanings, which is ideal to test the robustness of planet b's orbital solution. We used the quasi-periodic kernel implemented by \code{celerite} \citep{Foreman-Mackey2017} as offered in \code{RadVel}. The results for our GP $+$ planet b model are shown in Figure~\ref{fig:784one_planet_fit_GP} and Table~\ref{tab:priors-one-planet-fit-GP}. The derived $M_b\sin{i_b}$ is $9.49^{+0.97}_{-0.94}\ \rm{M_{\oplus}}$, consistent with our results in Table~\ref{tab:planet-pars} and also consistent with the results using the model with the two linear trends.

\begin{table*}
 	\centering
 	\caption{Priors and best-fit values of one-planet fit}
 	\label{tab:priors-one-planet-fit}
 	\begin{tabular}{llcr} 
 		\hline
 		\hline
		Parameter & Prior & Best-fit & Description\\
 		\hline
		$P_b$~(day) & $\mathcal{N}(2.7970364, 0.00001^2)$ & $2.797036^{+0.000010}_{-0.000010}$ & Orbital period of TOI-784b\\
		$t_{0,b}$~\rev{(BJD)} & $\mathcal{N}(2459336.61220, 0.001^2)$ & $2459336.6122^{+0.0010}_{-0.0010}$ & Time of transit-center for TOI-784b\\
		$\sqrt{e_b}\sin\omega_b$ & Fixed & 0\\
		$\sqrt{e_b}\cos\omega_b$ & Fixed & 0\\
		$K_b\ (\rm{m/s})$ & $\mathcal{U}(-20.0, 20.0)$ & $4.59^{+0.28}_{-0.29}$ & RV semi-amplitude of TOI-784b\\
		$\dot{\gamma}_1$ & \code{RadVel} default & $0.643^{+0.067}_{-0.065}$ & Linear trend term in the RV fit\\
		$\ddot{\gamma}_1$ & Fixed & 0 & Curvature term in the RV fit\\
		$\dot{\gamma}_2$ & \code{RadVel} default & $0.59^{+0.18}_{-0.18}$ & Linear trend term in the RV fit\\
		$\ddot{\gamma}_2$ & Fixed & 0 & Curvature term in the RV fit\\
		$\mu_{\rm{PFS},1}$ & $\mathcal{N}(0.0, 10.0^2)$ & $0.50^{+0.23}_{-0.23}$ & Velocity zero-point for PFS\\
		$\mu_{\rm{PFS},2}$ & $\mathcal{N}(0.0, 10.0^2)$ & $-2.12^{+0.44}_{-0.43}$ & Velocity zero-point for PFS\\
		$\sigma_{\rm{PFS}}$ & $\mathcal{U}(0.0, 15.0)$ & $0.77^{+0.24}_{-0.23}$ & Jitter term for PFS\\
 		\hline
 	\end{tabular}
\end{table*}

\begin{table*}
 	\centering
 	\caption{Priors and best-fit values of one-planet fit including a GP model}
 	\label{tab:priors-one-planet-fit-GP}
 	\begin{tabular}{llcr} 
 		\hline
 		\hline
		Parameter & Prior & Best-fit & Description\\
 		\hline
 		Planetary parameters\\
		$P_b$~(day) & $\mathcal{N}(2.7970364, 0.00001^2)$ & $2.797036^{+1.0e-5}_{-1.1e-5}$ & Orbital period of TOI-784b\\
		$t_{0,b}$~\rev{(BJD)} & $\mathcal{N}(2459336.61220, 0.001^2)$ & $2459336.6122^{+0.0011}_{-0.0010}$ & Time of transit-center for TOI-784b\\
		$\sqrt{e_b}\sin\omega_b$ & Fixed & 0\\
		$\sqrt{e_b}\cos\omega_b$ & Fixed & 0\\
		$K_b\ (\rm{m/s})$ & $\mathcal{U}(-20.0, 20.0)$ & $4.59^{+0.32}_{-0.30}$ & RV semi-amplitude of TOI-784b\\
		$\dot{\gamma}$ & Fixed & 0 & Linear trend term in the RV fit\\
		$\ddot{\gamma}$ & Fixed & 0 & Curvature term in the RV fit\\
		$\mu_{\rm{PFS}}$ & $\mathcal{N}(0.0, 10.0^2)$ & $-0.30^{+6.0}_{-4.8}$ & Velocity zero-point for PFS\\
		$\sigma_{\rm{PFS}}$ & $\mathcal{U}(0.0, 15.0)$ & $0.82^{+0.25}_{-0.22}$ & Jitter term for PFS\\
		Gaussian process parameters\\
		$B_{\rm{PFS}}$ & $\mathcal{U}(0.0001, 10000)$ & $76^{+680}_{-61}$ & GP hyperparameter\\
        $C_{\rm{PFS}}$ & $\mathcal{U}(0.0001, 10000)$ & $0.006^{+0.23}_{-0.0057}$ & GP hyperparameter\\
        $L_{\rm{PFS}}$ & $\mathcal{U}(0.0001, 1.0e+7)$ & $6254^{+2600}_{-3300}$ & GP hyperparameter\\
        $P_{\rm rot, \rm PFS}$ & $\mathcal{U}(10.0, 450.0)$ & $54^{+160}_{-27}$ & GP hyperparameter\\
 		\hline
 	\end{tabular}
\end{table*}

\begin{figure*}
    \centering
	\includegraphics[width=160mm]{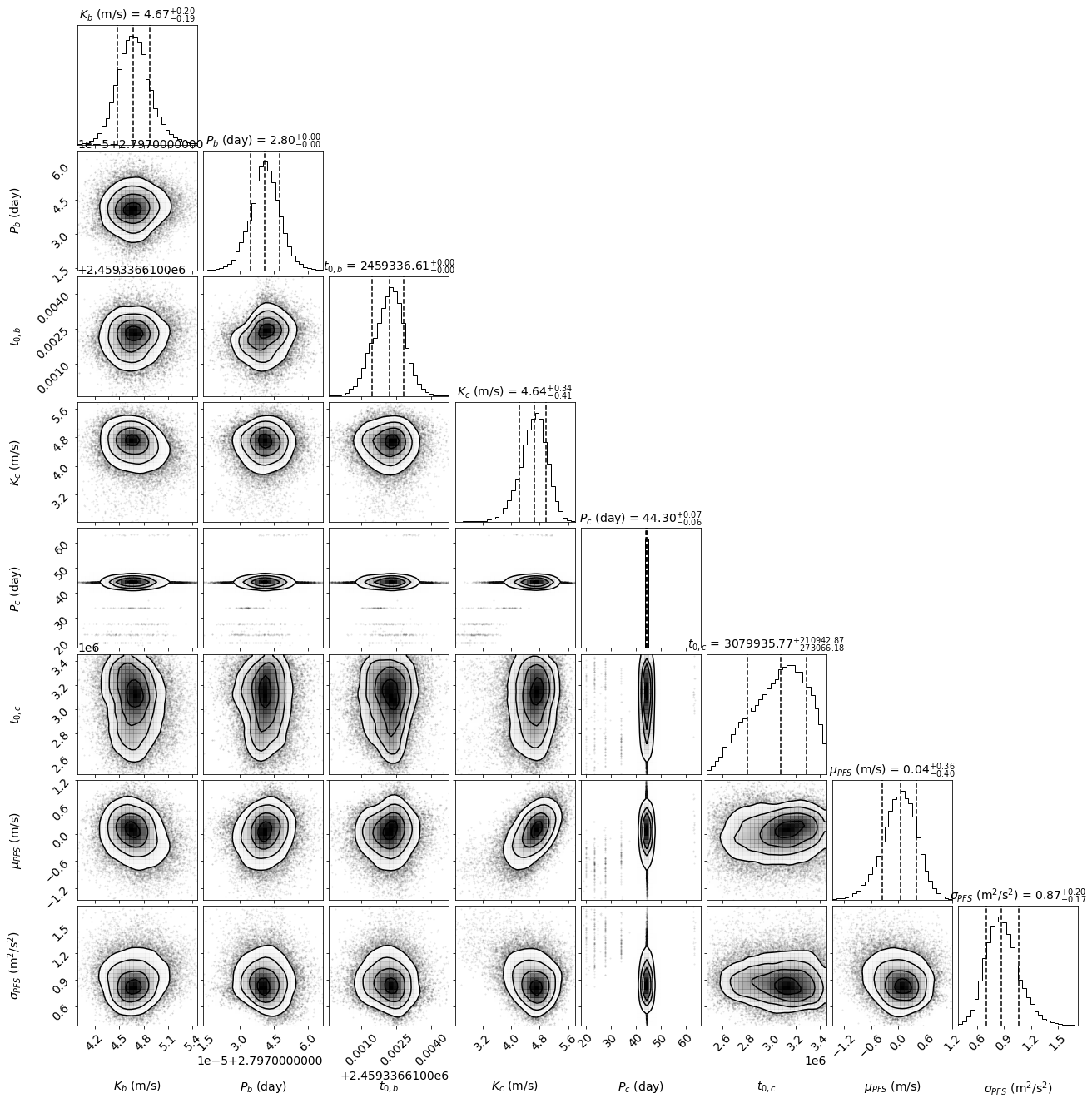}
    \caption{Corner plot of all fitted parameters in the two-planet fit using \code{Juliet}. See Section~\ref{sec:two-planet-fit-juliet} for more details.}
    \label{fig:784_2pl_fullcorner_juliet}
\end{figure*}




\end{document}